\documentclass[twocolumn,onecolappendix,tighten,times]{aastex62}

\usepackage{tikz}
\usepackage{amsmath}
\usepackage{multirow}
\usepackage{etoolbox}

\usepackage{nicefrac}
\usepackage{enumitem}
\usepackage{pbox}
\usepackage{soul}
\usepackage{float}
\usepackage{lineno}

\let\OLDthebibliography\thebibliography
\renewcommand\thebibliography[1]{
  \OLDthebibliography{#1}
  \setlength{\parskip}{-0.5pt}
  \setlength{\itemsep}{-0.5pt plus 0.3ex}
}

\newcommand*\mycirc[1]{%
   \begin{tikzpicture}
     \node[draw,circle,inner sep=0pt, minimum size=0.5pt] {#1};
   \end{tikzpicture}}
   
   \makeatletter
\newcommand*{\mySpecialfootnotes}[1]{%
  \patchcmd{\@footnotetext}{\floatingpenalty\@MM}{\floatingpenalty#1\relax}%
           {}{\errmessage{Couldn't patch \string\@footnotetext}}%
}

\shorttitle{Atmosphere Origins for Exoplanet sub-Neptunes}
\shortauthors{Kite et al.}

\begin{document}


\title{\Large{\textbf{Atmosphere Origins for Exoplanet Sub-Neptunes}}}%



\author{Edwin S. Kite}
\affil{Department of the Geophysical Sciences, University of Chicago, Chicago, IL (kite@uchicago.edu)}

\author{Bruce Fegley Jr.}
\affil{Planetary Chemistry Laboratory, McDonnell Center for the Space Sciences \& Department of Earth \& Planetary Sciences, \\ Washington University, St Louis, MO}

\author{Laura Schaefer}
\affil{School of Earth Sciences, Stanford University, Palo Alto, CA}
\author{Eric B. Ford}\affiliation{Department of Astronomy and Astrophysics, The Pennsylvania State University, University Park, PA}
\affiliation{Center for Exoplanets and Habitable Worlds, The Pennsylvania State University, University Park, PA}
\affiliation{Institute for CyberScience}
\affiliation{Pennsylvania State Astrobiology Research Center}

\begin{abstract}
\noindent Planets with 2 $R_{\Earth}$ $<$ $R$ $<$ 3 $R_{\Earth}$ and orbital period~$<$100~d are abundant; these sub-Neptune exoplanets are not well understood. For example, \emph{Kepler} sub-Neptunes are likely to have deep magma oceans in contact with their atmospheres, but little is known about the effect of the magma on the atmosphere.  Here we study this effect using a basic model,  assuming that volatiles equilibrate with magma at $T$~$\sim$~3000~K. For our Fe-Mg-Si-O-H model system, we find that chemical reactions between the magma and the atmosphere and dissolution of volatiles into the magma are both important. Thus, magma matters. For H, most moles go into the magma, so the mass target for both H$_2$ accretion and H$_2$ loss models is weightier than is usually assumed. The known span of magma oxidation states can produce sub-Neptunes that have identical radius but with total volatile masses varying by 20-fold. Thus, planet radius is a proxy for atmospheric composition but not for total volatile content. This redox diversity degeneracy can be broken by measurements of atmosphere mean molecular weight. We emphasise H$_2$ supply by nebula gas, but also consider solid-derived H$_2$O. We find that adding H$_2$O to Fe probably cannot make enough H$_2$ to explain sub-Neptune radii because ~$>$10$^3$-km thick outgassed atmospheres have high mean molecular weight. The hypothesis of magma-atmosphere equilibration links observables such as atmosphere H$_2$O/H$_2$ ratio to magma FeO content and planet formation processes. Our model's accuracy is limited by the lack of experiments (lab and/or numerical) that are specific to sub-Neptunes; we~advocate for such experiments. 
 \end{abstract}
\vspace{0.1in}



\section{Introduction.}
\noindent

\begin{figure}
\centering
\includegraphics[width=1.00\columnwidth,clip=true,trim={0mm 0mm 0mm 0mm}]{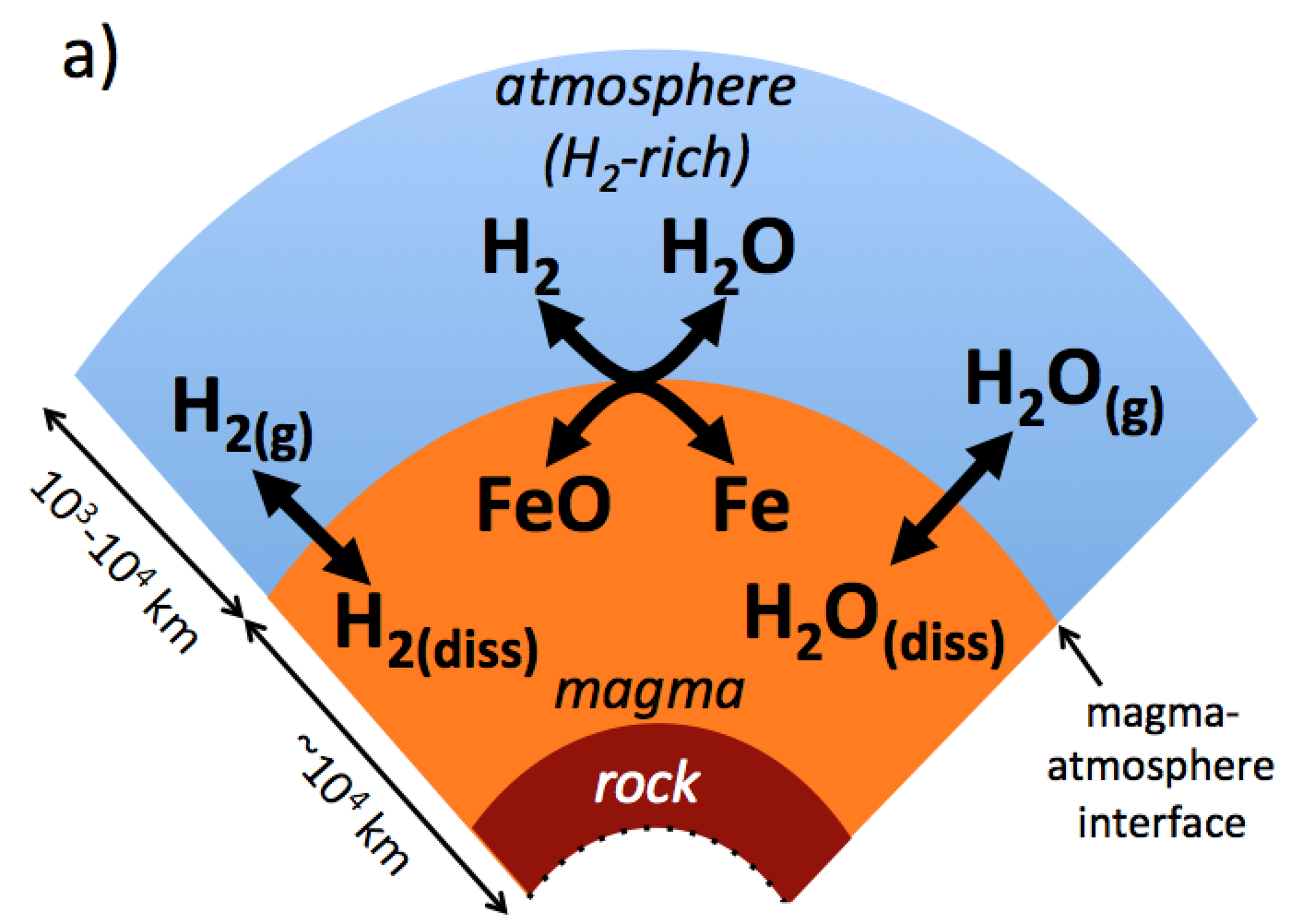}
\includegraphics[width=0.85\columnwidth,clip=true,trim={0mm 0mm 0mm 0mm}]{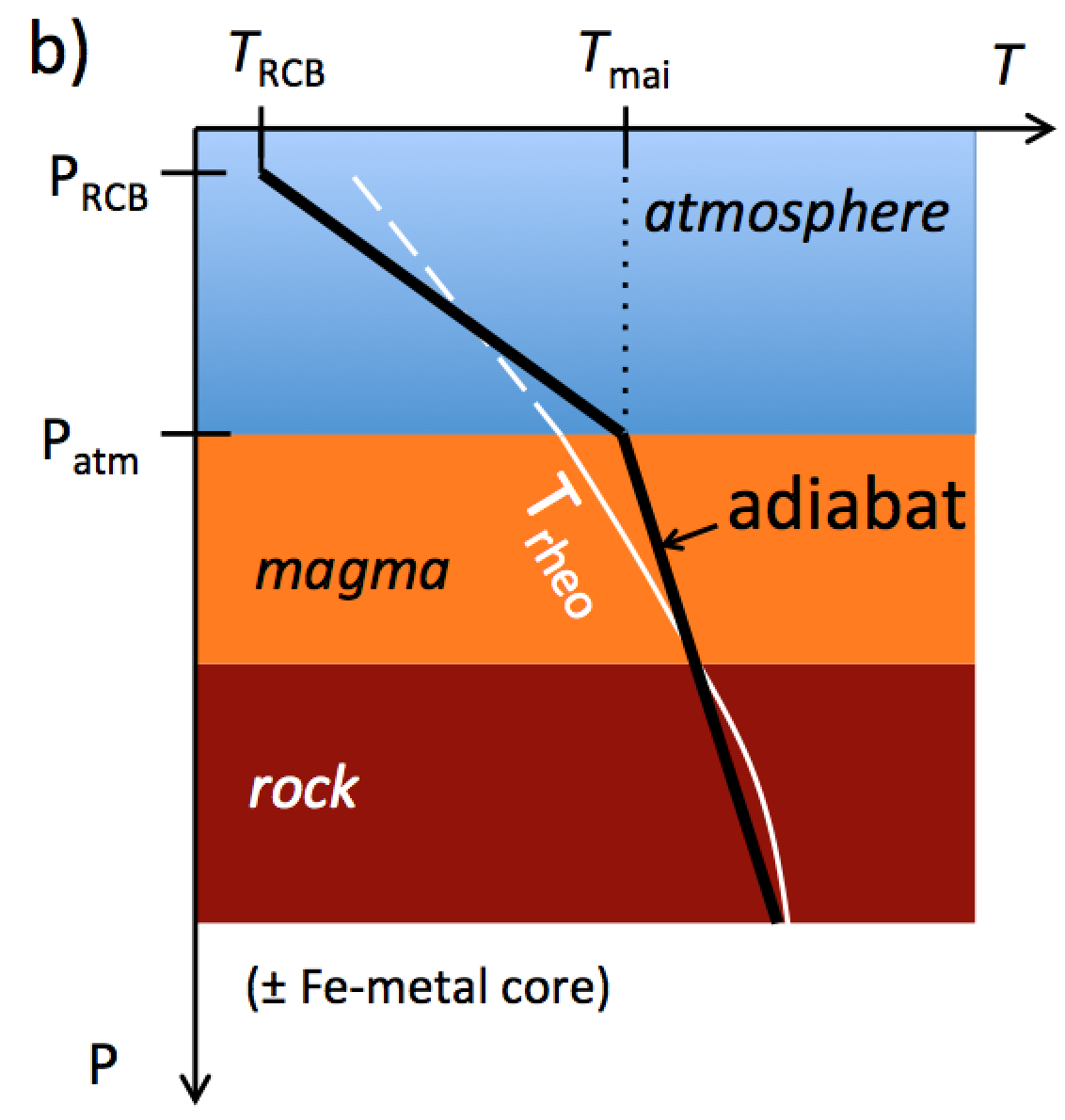}
\caption{(a) Structure of \emph{Kepler} sub-Neptunes assumed in this paper. For the first time, we model redox reactions between magma and atmosphere, specifically H$_2$+FeO $\leftrightarrow$ H$_2$O + Fe, on sub-Neptunes. (b) Temperature versus pressure plot, showing adiabats within the atmosphere and the magma (black line). $T_{rheo}$ (white line) corresponds to the temperature of the rheological transition ($\sim$40\% melt fraction) for rock. $RCB$ = radiative-convective boundary. $T_{mai}$ = temperature at the magma-atmosphere interface. For real sub-Neptunes the equilibrium temperature ($T_{eq}$), the effective temperature ($T_{eff}$), and the temperature at the $RCB$ ($T_{RCB}$), are related by $T_{eq}$ $<$ $T_{eff}$ $\lesssim$ $T_{RCB}$. In this paper we make the approximation that these three temperatures are equal. }  
\label{fig:pieslice}
\end{figure}

\noindent 
Using exoplanet atmosphere data to constrain planet formation and evolution is a core goal of exoplanet research \citep{Charbonneau2018}. So far, most data has come from exo-Jupiters. However,  smaller-radius worlds are far more intrinsically common (e.g., \citealt{Hsu2019}), and the characterization of their atmospheres is underway (e.g., \citealt{Fraine2014,Knutson2014,Wakeford2017,Morley2017,Benneke2019a}). At $R$~$<$~4~$R_{\Earth}$, most confirmed exoplanets are sub-Neptunes: worlds with $R~=~1.6-3.2$~$R_{\Earth}$ and density $<$~4~g/cc (e.g., \citealt{Fulton2017, Rogers2015,Wolfgang2016}). Sub-Neptunes probably have 10$^{3}$-10$^4$-km-deep H$_2$-rich atmospheres cloaking rocky cores (e.g., \citealt{OwenWu2017,vanEylen2018}), at least for orbital period~$<$~100~d. This implies that sub-Neptunes are mostly atmosphere by volume, and mostly silicate by mass (Fig.~\ref{fig:pieslice}). Thus we might expect silicate-atmosphere interactions would set atmosphere mass and composition. Two (coupled) interactions matter most (Fig.~\ref{fig:pieslice}): dissolution of the atmosphere into the magma, and redox reactions involving atmosphere and magma (e.g., \citealt{Hirschmann2012,Schaefer2016,ChachanStevenson2018}). Surprisingly, however, no previous study has investigated how these processes set atmosphere composition for sub-Neptunes. 

   \begin{figure*}[th]
\centering
\includegraphics[width=1.9\columnwidth,trim={0mm 0mm 0mm 0mm}]{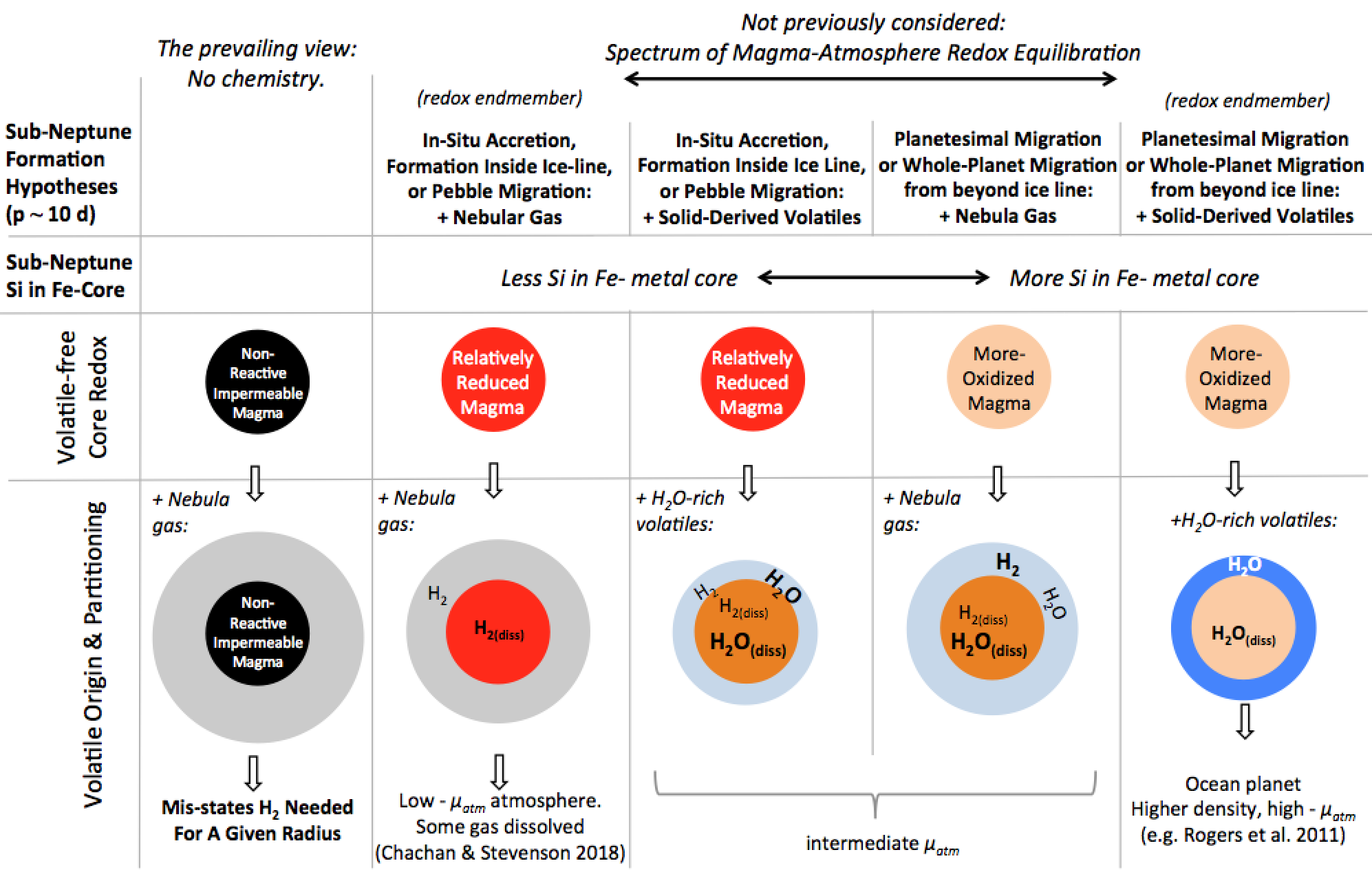}
\caption{Magma-atmosphere interactions considered in this paper. ($\mu_{atm}$ = atmospheric mean molecular weight; (diss.) = volatiles dissolved in magma. Magma oxidation may result from either net oxidation of the planet's materials, or dissolution of Si into the Fe-metal core (both processes might contribute; Appendix~B). }  
\label{fig:cartoon}
\end{figure*}

 \emph{What is the volatile content of sub-Neptunes, where did it come from, and where is it today?}  The most important volatile element is H. H on sub-Neptunes is stored as H$_2$ in the atmosphere; as H$_2$O in the atmosphere; as H$_2$ contained within silicate (magma or rock) \citep{ChachanStevenson2018}; as H$_2$O dissolved in the silicate; and perhaps as H dissolved in Fe-metal (e.g., \citealt{Clesi2018}). So far, sub-Neptune formation models have emphasized H~stored in the atmosphere, and emphasized H sourced from the nebula. This understates the H needed to produce an atmosphere of a given mass, neglects the possibility of H$_2$ generation via Fe-oxidation \citep{Rogers2011}, and ignores H$_2$O generation by reduction of FeO \citep{Sasaki1990}. New data from contaminated white dwarfs indicate extrasolar silicates with high FeO content \citep{Doyle2019}; reaction between such silicates and nebula gas would generate H$_2$O. Moreover, magma redox can probe planet formation (e.g. \citealt{Urey1952,Wanke1981}), and so atmospheric constraints on magma-atmosphere reactions and thus magma redox can probe planet formation (Fig.~\ref{fig:cartoon}).

 We believe this is the first work on magma-atmosphere reaction for sub-Neptunes (for an overview of related Solar System work, see Appendix E). Of previous related studies -- e.g. \citet{IkomaGenda2006, RogersSeager2010, Rogers2011,Schaefer2016, Massol2016, SchaeferFegley2017, Ikoma2018} -- the closest in intent to our own is that of \citet{ElkinsTantonSeager2008a}. Their study uses elemental mass balance to set upper limits on volatile abundance, but does not attempt to solve for redox equilibrium. \citet{ChachanStevenson2018} use an ideal-gas model to study the solubility of H$_2$ in magma -- the H$_2$ redox endmember in Fig.~\ref{fig:cartoon}.  In \citet{Kite2019}, we used a real-gas H$_2$ model to also study this end-member case, finding qualitatively different results from the ideal-gas model. 
 
Because this is the first study of atmosphere-magma reaction for sub-Neptunes, in the remainder of this paper, we use a basic model (\S2). We neglect photochemistry and we neglect fractionating escape-to-space (retention of O / H$_2$O versus loss of H). Results are given in \S3. We emphasize applications and tests in our analysis (\S4). We discuss in \S5 and conclude in \S6. Our main results are in Figs.~\ref{fig:5outcomes}-\ref{fig:muboost}.

 \subsection{Most \emph{Kepler} Sub-Neptunes Have Massive Magma Oceans In Direct Contact With The Atmosphere}

\begin{figure}
\centering
\includegraphics[width=1.05\columnwidth,clip=true,trim={2mm 0mm 0mm 0mm}]{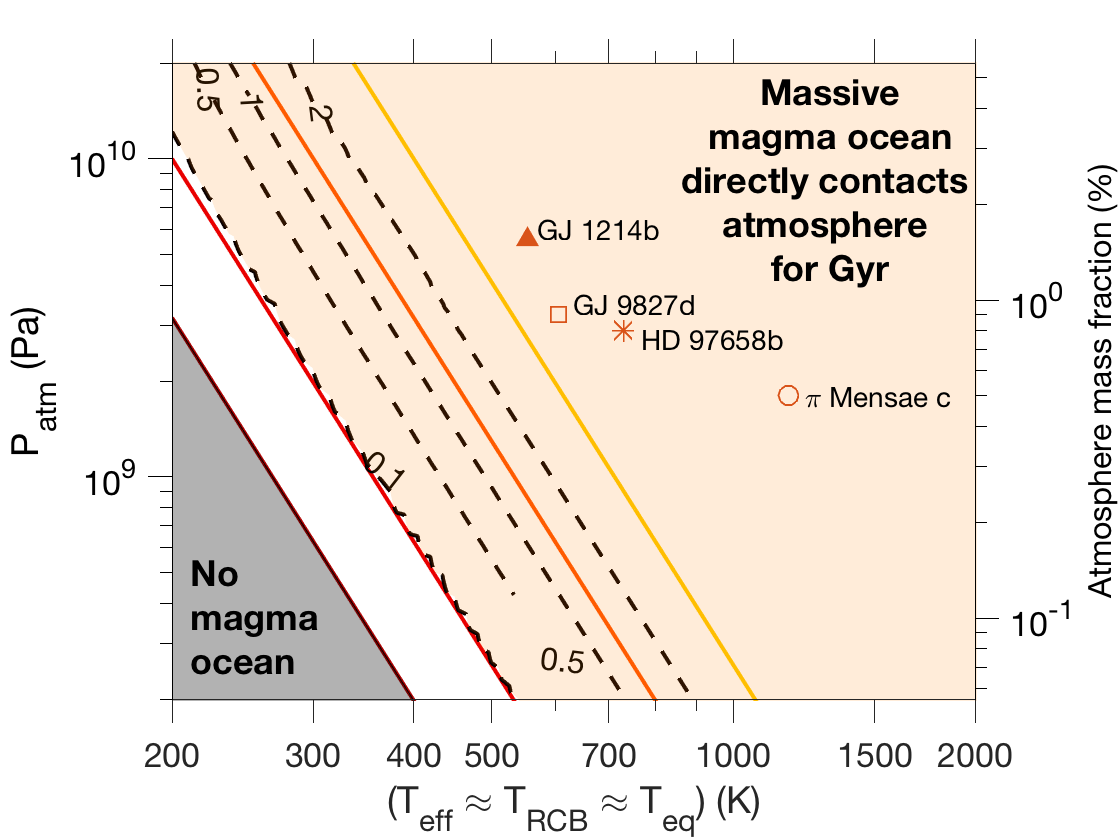}
\caption{How magma ocean mass increases with atmospheric thickness. Output from a toy model of sub-Neptune thermal structure (Appendix~A). Dashed lines correspond to magma ocean mass, labeled in Earth-masses of magma, for a volatile-free planet mass of 5~$M_\Earth$. Colored lines correspond to temperatures at the magma-atmosphere interface of (going from left to right) 1500~K, 2000~K, 3000~K, and 4000~K. Magma ocean masses in excess of 2 Earth masses are not plotted because this corresponds to a magma $P$ range that is little-explored by experiment. Planets of varying masses are over-plotted: GJ~1214b (parameters from \citealt{Nettelmann2011}), GJ~9827d \citep{Rice2019}, HD~97658b (\citealt{Dragomir2013} / \citealt{VanGrootel2014}; upper limit on atmosphere mass), and $\pi$~Mensae~c \citep{Huang2018}.}  
\label{fig:howmuchmagma}
\end{figure}

\noindent Models show that an Earth-composition planet will double in radius if it gathers $\sim$3\% of its own weight into a nebula-composition atmosphere (for orbital period~$\sim$10~d) (e.g.,~\citealt{LopezFortney2014,Bodenheimer2018}). However, these models treat the remaining $\sim$97\% of the planet's mass (silicates plus Fe-metal) as chemically inert. To the contrary, the magma-atmosphere interface on sub-Neptunes has a temperature ($T_{mai}$) hotter than the silicate liquidus, and the interface is chemically re-active and permeable  \citep{SchaeferElkinsTanton2018}. $T_{mai}$~is~hot because \emph{Kepler} sub-Neptunes form hot, and the atmosphere insulates the silicates. Sub-Neptune $T_{mai}$ is much hotter than sub-Neptune photosphere temperature $T_{eff}$. Pressure at the magma-atmosphere interface is 1-10~GPa (Fig.~\ref{fig:howmuchmagma}).
 
The magma ocean forms a global shell underneath the atmosphere (Fig.~\ref{fig:pieslice}). This shell is most massive for young planets with thick atmospheres that are close to the star. $T_{mai}$ versus time can be tracked with complex models (e.g. \citealt{HoweBurrows2015,ChenRogers2016,Bodenheimer2018, Vazan2018a}). However, these studies all present $T_{mai}$ for only a few cases and with the exception of \citet{Vazan2018a} they do not estimate magma ocean mass. We wanted to build intuition for how magma ocean mass depends on $T_{eff}$ and $P_{atm}$. Therefore we wrote a toy model of sub-Neptune thermal structure (Appendix~A). The toy model output is shown in Fig.~\ref{fig:howmuchmagma}. The toy model indicates that for atmosphere mass down to 0.2~wt\%, extensive magma occurs for $T_{eff}$~$>$~400~K. 

Above the liquidus, silicate magma is runnier than water, so if the magma layer convects, then the magma and  atmosphere can stay equilibrated  \citep{Massol2016}. In this paper, we assume full equilibration between the atmosphere and fully molten silicates. For this assumption to be correct, a neccessary condition is that the silicates are fully molten. This condition is more likely to be satisfied for worlds that are young, retain thick atmospheres, and are in short-period orbits (high $T_{eq}$).

Magma is chemically potent. For example, the magma can be the host of most of the planet's exchangeable electrons (i.e., the magma can dominate the redox budget). This corresponds to planets with $O$(1~wt\%) atmosphere mass and $O$(5~wt\%) total volatile mass. We focus on this case here. 

\section{Method.}

\noindent To explore magma effects on sub-Neptune atmospheres, we assume atmosphere equilibrates with a well-stirred magma ocean. Our model makes the following simplifications:- (a)~We~consider only the elements Fe, Mg, Si, O, and H. Chemically reduced carbon compounds may also contain H; for simplicity we omit consideration of them here. We also restrict ourselves to the range of magma elemental compositions for which SiO$_2$ is a major constituent. (b)~We set 2000~K~$\le$~$T_{mai}$~$\le$~3000~K, so that the magma-atmosphere interface is molten but the vapor pressure of the magma is small relative to the total atmospheric pressure \citep{Fegley2016,SossiFegley2018}. This $T_{mai}$ is at the low end of the $T_{mai}$ output by thermal evolution models for multi-Gyr-old sub-Neptunes (e.g.~\citealt{BodenheimerLissauer2014,HoweBurrows2015,Vazan2018b}). We consider lower $T_{mai}$ in \S4.4.  (c)~We~neglect Fe$^{3+}$ (i.e.,~Fe$_2$O$_3$). We expect Fe$^{3+}$ will be a minor constituent of a magma ocean equilibrated with a H$_2$-dominated atmosphere. Equal thermodynamic activities of FeO and Fe$_2$O$_3$ in magma at 3000~K require \emph{f}O$_2$ = 28 bars, much higher than expected from thermal dissociation of steam at any $P$ and $T$ below 1 kilobar and 3500~K.
(d)~We assume that metal (if present) is pure Fe for these calculations; in reality, the metal will be an Fe-dominated alloy. We may be (slightly) underestimating \emph{f}O$_2$ by doing this. (e)~We track non-ideal behavior of both H$_2$ and H$_2$O  (Appendix~C), but we assume ideal mixing of H$_2$ and H$_2$O. This is a valid assumption both under mineral-free conditions in the atmosphere for $T$~$>$~650K, and also at the magma-atmosphere interface given our model assumptions  \citep{SewardFranck1981,Bali2013,SoubiranMilitzer2015}. We neglect joint-solubility effects. (f)~We~use a single value of gravitational acceleration $g$, corresponding to 1.2$\times$ the bare-rock radius, to convert from bottom-of-atmosphere pressure to atmosphere column mass. (g)~We~neglect the effect of dissolved volatiles on core mass. 

Consider a well-stirred magma ocean that is redox-buffered by coexistence of liquid Fe-metal and FeO-bearing magma:

\begin{equation}
2\,\mathrm{Fe}_{\mathrm{(liq)}} +  \,\mathrm{O}_{2\mathrm{(g)}} = 2\,\mathrm{Fe}\mathrm{O_{\mathrm{(liq)}}}
\label{eqn:feo2}
\end{equation}

\noindent The equilibrium constant for Reaction \ref{eqn:feo2} is 

\begin{equation}
K_1 = \frac{[\mathrm{FeO}]^2}{[\mathrm{Fe}]^2 f\mathrm{O_2}} = \frac{[\mathrm{FeO}]^2}{f\mathrm{O_2}} \mathrm{\,for \,[Fe] = 1}
\end{equation}

\begin{figure}
\centering
\includegraphics[width=0.9\columnwidth,clip=true,trim={0mm 0mm 0mm 0mm}]{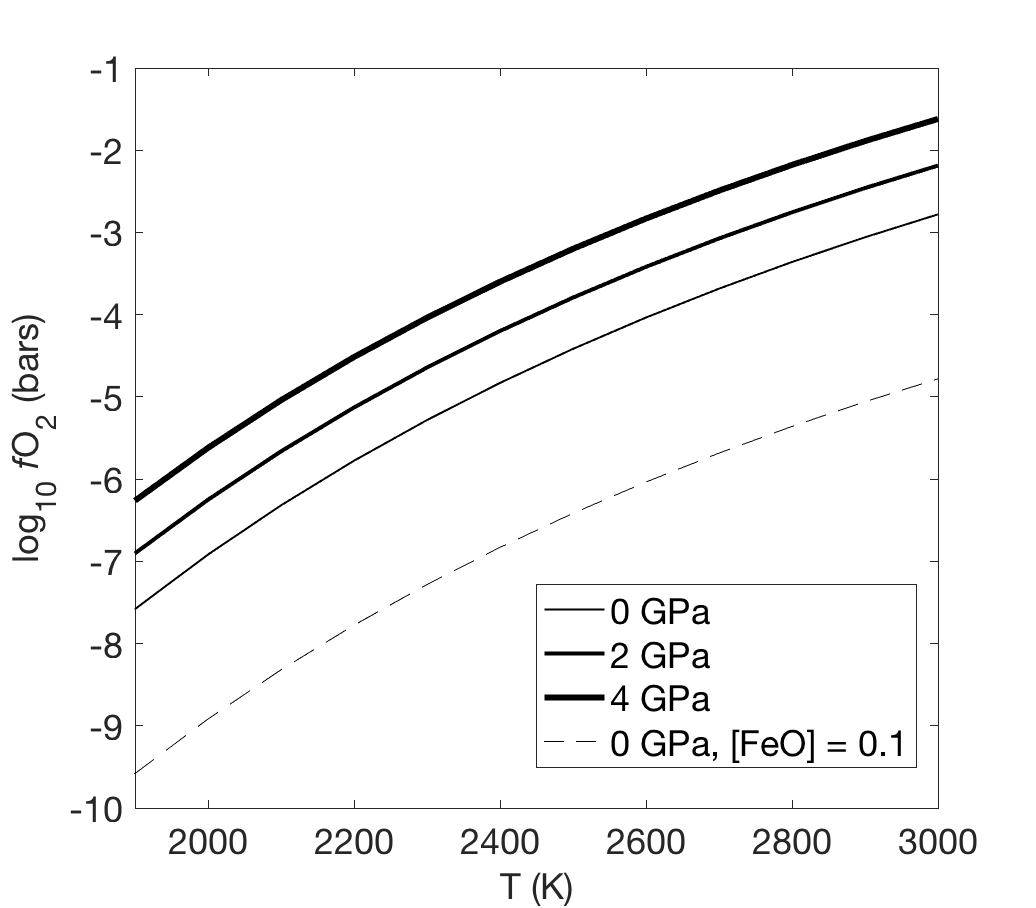}
\caption{The oxygen fugacity, \emph{f}O$_2$, corresponding to coexistence of liquid Fe-metal and liquid FeO. The solid lines show the results for pure liquid FeO, and the dashed line shows the 100-fold reduction in \emph{f}O$_2$ that results from reducing the concentration of FeO in the silicate magma to 10\%.}  
\label{fig:fo2}
\end{figure}

\noindent The oxygen fugacity \emph{f}O$_2$ corresponding to coexistence of liquid Fe-metal and liquid FeO is a function of $T$, $P$, and the chemical activity of FeO in the magma (Fig.~\ref{fig:fo2}). Fugacity is a measure of chemical reactivity, expressed in units of pressure. Fugacity is 1-10$\times$ the partial pressure of the gas for the range of conditions we consider (Appendix~C). We obtain \emph{f}O$_2$($T$, $P$) by~combining data for the FeO liquid equation of state of \citet{Armstrong2019} and for the Fe~liquid equation of state from \citet{Komabayashi2014}, assuming an activity coeefficient for FeO ($\gamma_{FeO}$) of 1.5  \citep{Holzheid1997}. The Gibbs free energy at 1~bar for Reaction~\ref{eqn:feo2} is obtained from the data in \citet{KowalskiSpencer1995}.

The resulting \emph{f}O$_2$$(T, P)$ for the Fe/FeO buffer is shown in Fig.~\ref{fig:fo2}. Both the \emph{f}O$_2$, and the corresponding O$_2$ partial pressure, are tiny compared to total atmospheric pressure (Fig.~\ref{fig:fo2}). Thus, \emph{f}O$_2$ is used in our model solely as a convenient book-keeping variable for redox. The dominant atmospheric species in our model are H$_2$ and H$_2$O.  

\emph{f}O$_2$ $\propto$ [FeO]$^2$, where [FeO] is the activity of FeO in magma (e.g., \citealt{Frost2008}). Moreover, \emph{f}O$_2$ is directly related to the H$_2$/H$_2$O ratio in the atmosphere \citep{Fegley2013}. The connection is the reaction 

\begin{equation}
2\mathrm{H}_{2\mathrm{(g)}}~+~\mathrm{O}_{2\mathrm{(g)}}~=~2\mathrm{H_2O_{(g)}}
\label{eqn:h2o2}
\end{equation}

\noindent The equilibrium constant for Reaction~\ref{eqn:h2o2} is 

\begin{equation}
K_3 = \left( \frac{f\mathrm{H_2O}}{f\mathrm{H_2}} \right) \frac{1}{ f\mathrm{O_2}}
\end{equation}

\noindent Using the equality of oxygen fugacity for Reactions~\ref{eqn:feo2}~and~\ref{eqn:h2o2}, substituting and rearranging then gives

\begin{equation}
\frac{[\mathrm{FeO}]^2}{K_1} = \frac{1}{K_3}\left(\frac{f\mathrm{H_2O}}{f\mathrm{H_2}}\right)^2
\end{equation}


\noindent where $f_i = \phi_i P_i$ = $\phi_i X_i P_{total}$, with $\phi_i$ a ``fugacity coefficient'' (Appendix~C), and $P_i$ = $X_i P_{total}$ is the~partial~pressure of the species with $X_i$ the mole fraction; and

\begin{equation}
K_{3} = \mathrm{exp}\left(\frac{-\Delta G^\mathrm{\circ}}{R \, T}\right);
\end{equation}

\noindent $\Delta G^\mathrm{\circ}$ increases by $\sim$120~kJ/mol for a 1000~K rise in $T$. We use the~expression
 
\begin{multline}
\Delta G^\mathrm{\circ}(T) =-4.8716\times10^{5} + 94.261574 \,T \\
+ 9.9275922\times10^{-3} \,T^2  -1.87633188\times10^{-6}\, T^3 \\
 + 1.2446526\times10^{-10}\,T^4) 
\label{eqn:deltagh2o2}
\end{multline}

\noindent (from the IVTAN Tables, \citealt{Glushko1999}). The net reaction is the sum of Reaction~\ref{eqn:feo2} and Reaction~\ref{eqn:h2o2}: 

 \begin{equation}
\mathrm{FeO + H_2 = Fe + H_2O}
\label{eqn:feoh2}
\end{equation}

\noindent The equilibrium constant for Reaction~\ref{eqn:feoh2} is

\begin{equation}
K_8 = \frac{[\mathrm{Fe}]}{[\mathrm{FeO}] } \frac{f\mathrm{H_2O}}{f\mathrm{H_2}} = \frac{1}{[\mathrm{FeO}] } \frac{f\mathrm{H_2O}}{f\mathrm{H_2}}  \mathrm{\,for \,[Fe] = 1}
\end{equation}

\noindent The equilibrium constant $K_8$ is equal to

\begin{equation}
K_8 = \left( \frac{K_3}{K_1} \right)^{1/2} = \frac{1}{[\mathrm{FeO}] } \frac{f\mathrm{H_2O}}{f\mathrm{H_2}}  
\label{eqn:endofbasicmodel}
\end{equation}

\noindent This basic model (Eqns. \ref{eqn:feo2}-\ref{eqn:endofbasicmodel}) shows that the atmosphere has a ratio of water to hydrogen that is proportional to the FeO content of the underlying magma. Thus, provided that no water clouds form in the cool upper layers of the atmosphere, spectroscopic constraints on O/H in the cool upper layers of the atmosphere probe the composition of deep magma.

Neither the atmospheric pressure, nor the FeO~content of the magma, are free parameters -- they are results of magma-atmosphere equilibration. Therefore, we need to consider not just buffering of the atmosphere by the magma, but also the more general case of atmosphere-magma chemical coupling. Our approach to this is described below.

The H$_2$ solubility at the top of the magma layer is set to 

\begin{equation}
X_{\mathrm{H2}}~=~9.3~\times~10^{-12}\, f_\mathrm{H2} \, \mathrm{exp}(-T_{\mathrm{0}}/T_{mai})
\end{equation}

\noindent where $X_\mathrm{H2}$ is the mass fraction of H$_2$ in the melt, with $T_{\mathrm{0}}~=~4000$~K. This follows the estimated molten-average-rock solubility from \cite{Hirschmann2012} (i.e., their estimated peridotite solubility). $f_\mathrm{H2}$ is calculated from $P_\mathrm{H2}$ using the procedure given in Appendix~C. This solubility is $\sim$5$\times$ lower than was used by \cite{ChachanStevenson2018}, who used a solubility for molten basalt. Our proposed temperature dependence of the H$_2$ solubility follows \cite{ChachanStevenson2018}. There are no direct measurements of H$_2$~solubility in liquid magma in the 2000-3000~K range.

The H$_2$O solubility at the top of the magma layer is approximated as \citep{Schaefer2016}

\begin{equation}
\left(\frac{X_{\mathrm{H2O}}}{0.01}\right) =  \left(\frac{P_{\mathrm{H2O}}}{241.5 \,\mathrm{MPa} }\right)^{0.74}
\end{equation}

\noindent where $X_\mathrm{H2O}$ is the mass fraction of H$_2$O in the melt. This expression is a curve fit to the results of \citet{Papale1997} for basaltic melt; the high-pressure solubility of water in peridotite liquid is unknown. 

Equilibration at the top of the magma layer sets volatile abundance throughout the well-stirred magma. This is because the solubility of volatiles goes up with depth within the magma. Therefore, saturation equilibrium at the magma-atmosphere interface implies sub-saturated conditions at great depth (no bubbles, and constant mole fraction of H in magma). For a discussion of what happens if convection stalls, see \S4.4.

Equipped with these solubilities and equilibrium constants, we then solve for mass balance for H between H$_2$~in atmosphere, H$_2$ in magma, H$_2$O in atmosphere, and ``H$_2$O'' in magma. Our workflow is detailed in Appendix~D. 

 \begin{figure*}
    \centering
        \centering
   \textbf{(a)}     \includegraphics[width=1.0\columnwidth]{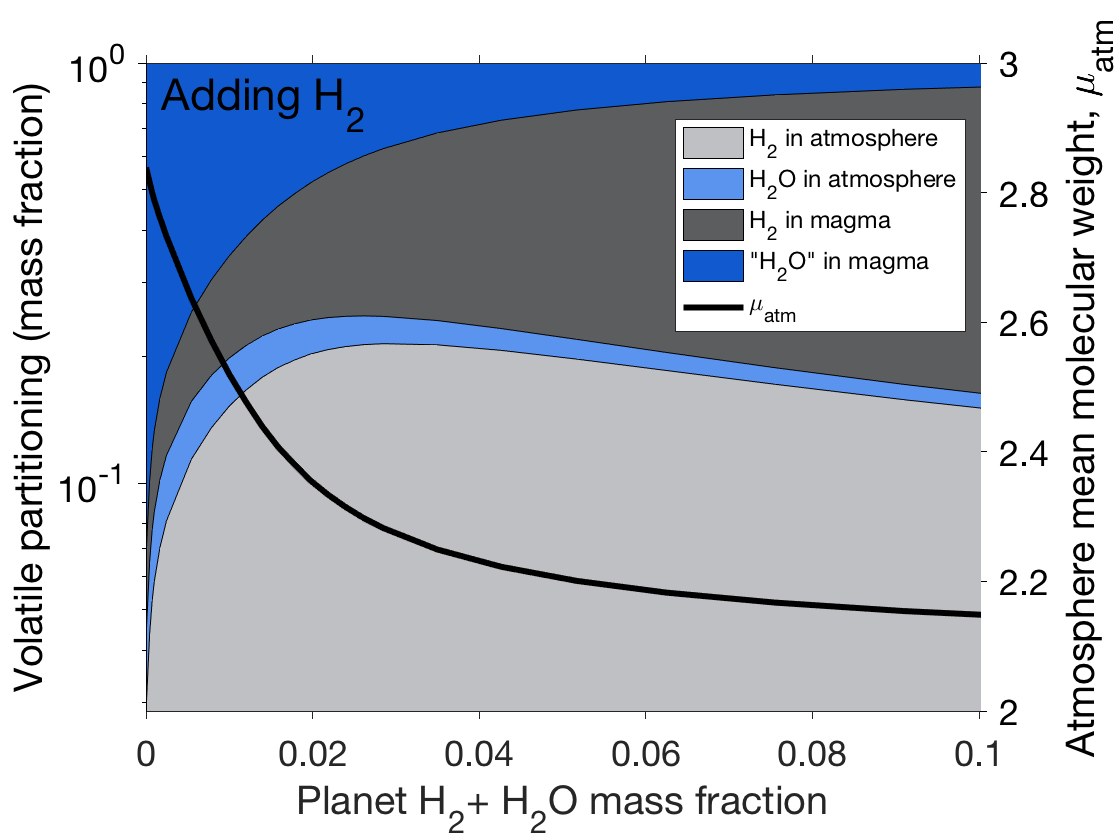}
        \centering
\textbf{(b)}        \includegraphics[width=1.0\columnwidth]{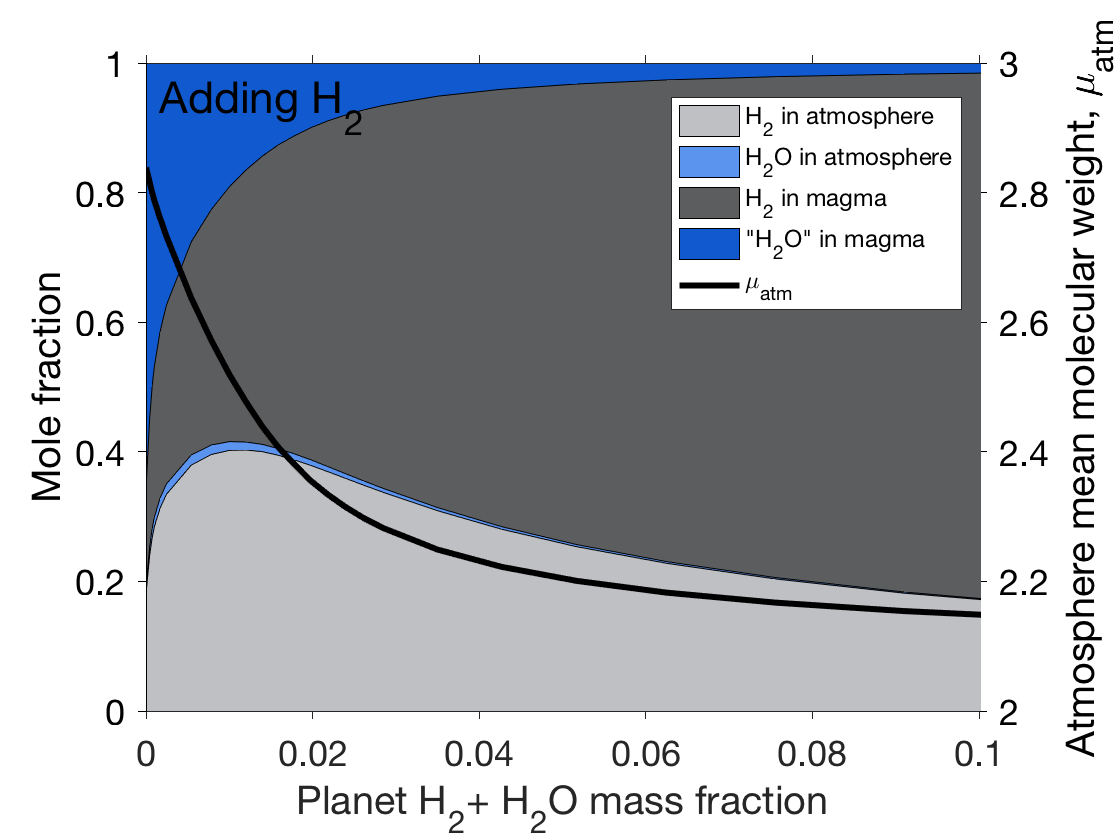}
        \centering
\textbf{(c)}        \includegraphics[width=1.0\columnwidth]{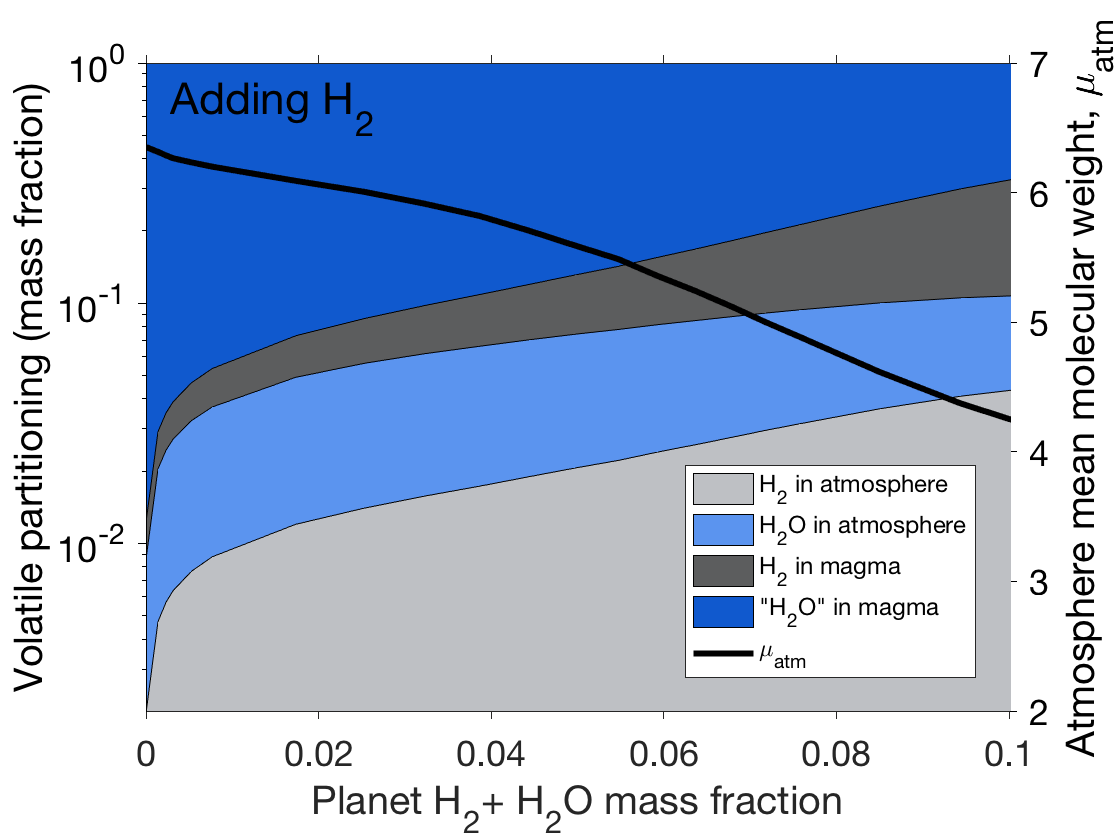}
        \centering
  \textbf{(d)}      \includegraphics[width=1.0\columnwidth]{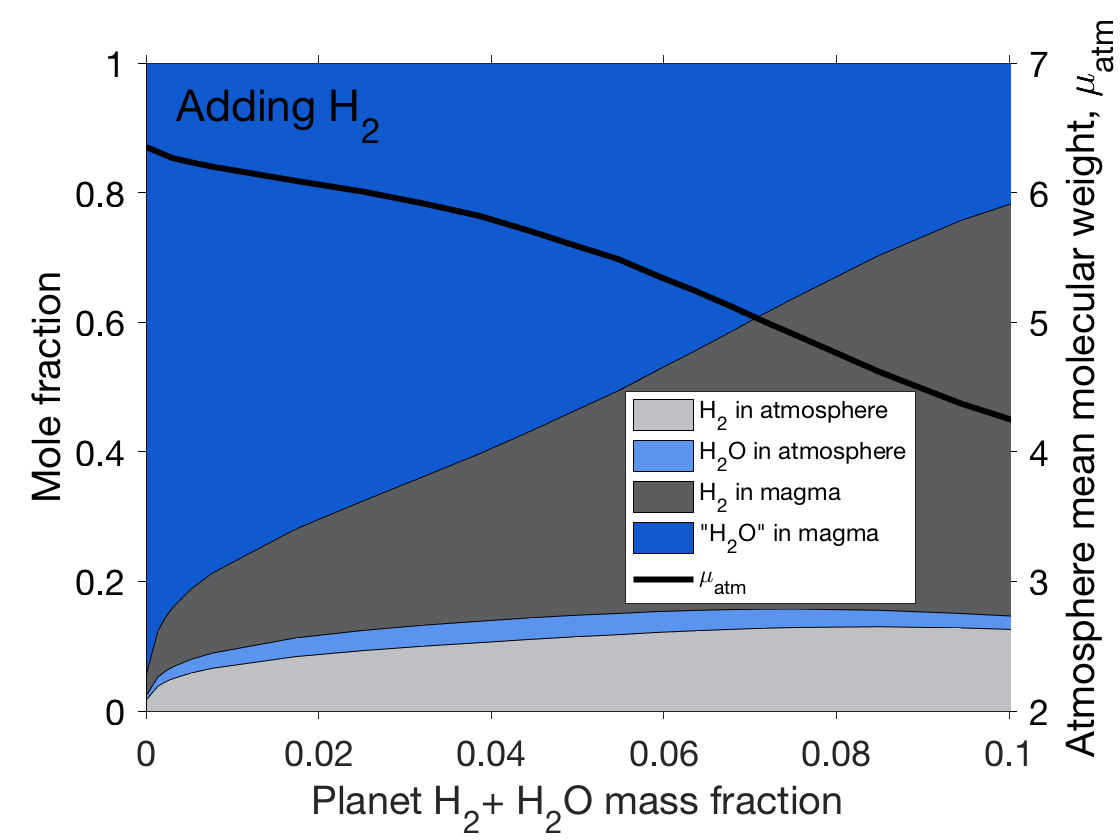}
\caption{How magma composition controls the fate of H: adding H$_2$ to a magma ocean of 3\nicefrac{1}{3}~Earth~masses (\nicefrac{2}{3} of the mass of a 5~Earth~mass planet) with a temperature at the  magma-atmosphere interface ($T_{mai}$) of 3000~K. Simplified model (see text). For details, see \S3.1.
\textbf{(a)} and \textbf{(b)} show the consequences of adding varying amount of nebular hydrogen to a planet with an initial Earth-like (8~wt\%) FeO content in its magma. (a) shows mass fractions and (b) shows mole fractions. 
\textbf{(c)} and \textbf{(d)} show the consequences of adding varying amount of nebular hydrogen to a planet with an initial FeO content in its magma similar to the mass of Fe-oxide that would be obtained from completely oxidizing Earth's mantle+core (49 wt\%). This gives an FeO content similar to the most oxidized of the extrasolar silicate oxidation states reported by \citet{Doyle2019}. (c) shows mass fractions and (d) shows mole fractions. }
    \label{fig:addh2results}
\end{figure*}

\vfill\null

\section{Results}

\subsection{Adding H$_2$ To an FeO-bearing Magma Ocean \textnormal{(Fig.~\ref{fig:addh2results})}.}

\noindent The top row in Fig.~\ref{fig:addh2results} shows the consequences of adding varying amounts of nebula hydrogen to a planet with an initial Earth-like FeO content in its magma. Given our assumption of magma-atmosphere equilibration (discussed further in \S5.2), the end results are the same if all of the nebula hydrogen is added after the core forms, or if the nebula hydrogen is added as the core is still growing. 

Moving to the right on the x-axis of the plots in Fig.~\ref{fig:addh2results} corresponds to adding volatiles to the planet. In Fig.~\ref{fig:addh2results} the volatile we choose to add is nebula gas. Adding nebular gas is equivalent, in our Fe-Mg-Si-O-H system, to adding pure H$_2$ (we~neglect O in nebula gas). As H$_2$ is added, H$_2$O is generated, via FeO + H$_2$ $\rightarrow$ Fe + H$_2$O \citep{IkomaGenda2006}. Because (at relatively modest atmospheric pressure) H$_2$O is much more soluble in magma than is H$_2$, the dominant reservoir for H is water dissolved in the magma. As~H$_2$ is added, FeO~goes to Fe, so \emph{f}O$_2$ goes down, and the atmospheric H$_2$O/H$_2$ ratio goes down (it is proportional to [FeO]) (Fig.~\ref{fig:addh2results}). Thus, atmospheric mean molecular weight decreases as H$_2$ is added. 
f
Most H goes into the magma. H$_2$O stored in the magma (dark blue bands in~Fig.~\ref{fig:addh2results}) can outweigh H$_2$ in the atmosphere (light gray bands in~Fig.~\ref{fig:addh2results}), even when the atmosphere is mostly H$_2$.  The importance of the magma as a volatile store is boosted by H$_2$ dissolution into the magma, corresponding to the dark gray bands in Fig.~\ref{fig:addh2results} \citep{ChachanStevenson2018}. The mass fraction of volatiles in the atmosphere peaks at $<$25~wt\% ($>$75~wt\% of volatiles are in the magma). It declines at higher $P_{atm}$ because H becomes very~soluble in magma at high $P$ \citep{Kite2019}.

The lower row of Fig.~\ref{fig:addh2results} shows the consequences of adding varying amounts of nebula hydrogen to a planet with an initial FeO mass fraction of 0.487 in its magma. This fraction corresponds to a planet for which Fe is initially entirely present as FeO. This planet does not have an Fe-metal core until H$_2$ is added \citep{ElkinsTantonSeager2008b}. Even for an atmosphere corresponding to 3~wt~\% total volatiles, our initial-FeO-mass-fraction = 0.487 world has a $\mu_{atm}$~$\sim$~6 atmosphere (Fig.~\ref{fig:addh2results}c). This is because much of the nebular-sourced H$_2$ is oxidized to H$_2$O, and most of this H$_2$O is sequestered in the mantle.

Fig.~\ref{fig:addh2results} shows that even modest magma oxidation has a big effect. We add nebula gas, but in the resulting planet, H$_2$O is very important (Fig.~\ref{fig:addh2results}). Adding H$_2$ to FeO makes H$_2$O, and H$_2$O is very soluble in magma. This is a redox-enabled hydrogen~pump.

Real worlds apparently sample the full range of stochiometrically possible silicate FeO contents. Within the Solar System, silicate mantle FeO contents range from $\le$0.04~wt\%~Fe$^{(2+)}$O for Mercury \citep{Nittler2017}, to $\sim$20~wt\%~Fe$^{(2+)}$O for Vesta and Mars. \citet{Doyle2019} report extrasolar silicate oxidation states from contaminated white dwarf spectra ($n$~=~6). The highest of their values is roughly equivalent to our full-oxidation calculation (Fig.~\ref{fig:addh2results}c).

 \begin{figure*}
    \centering
        \centering
       \textbf{(a)} \includegraphics[width=1.0\columnwidth]{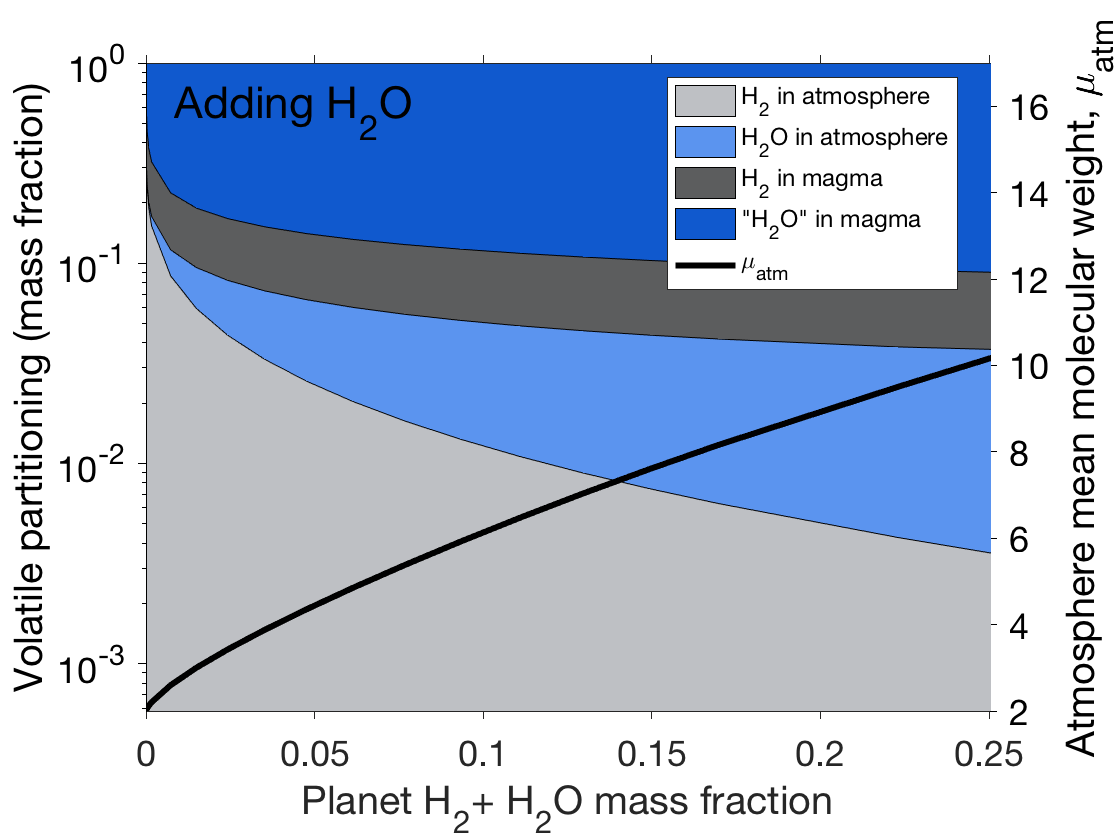}
        \centering
      \textbf{(b)}  \includegraphics[width=1.0\columnwidth]{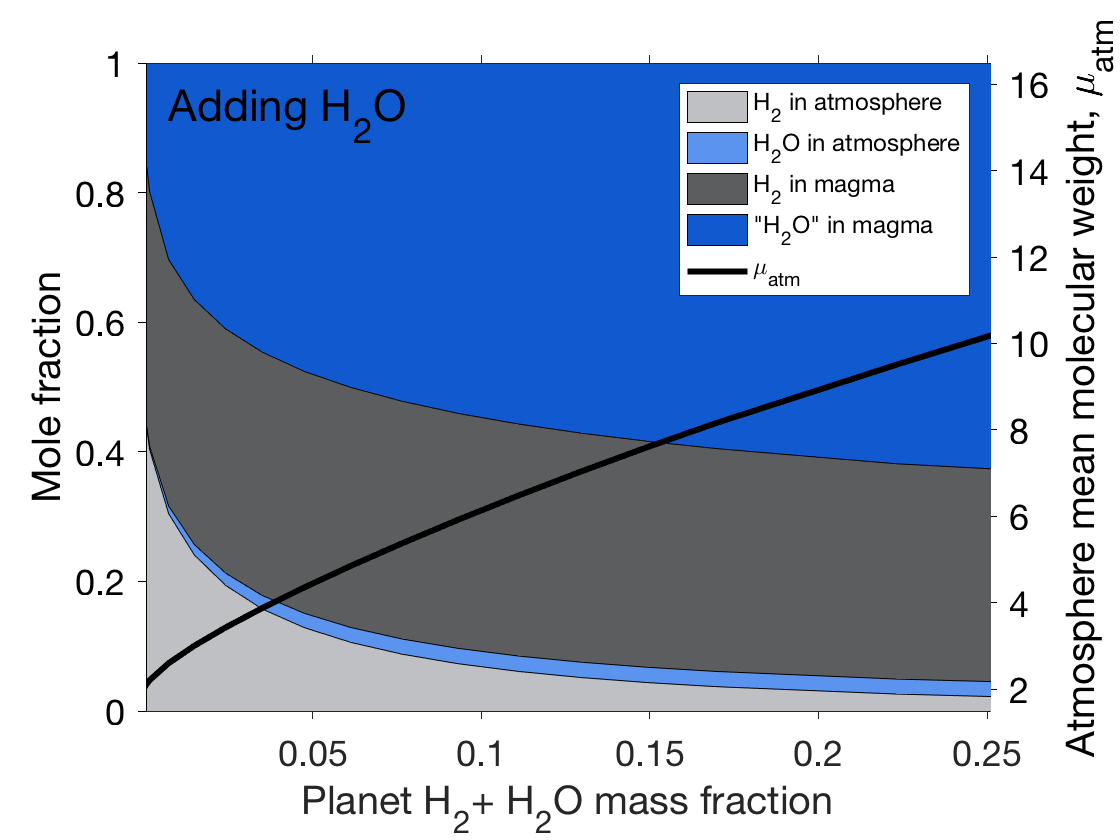}
                \centering
     \textbf{(c)}   \includegraphics[width=1.0\columnwidth]{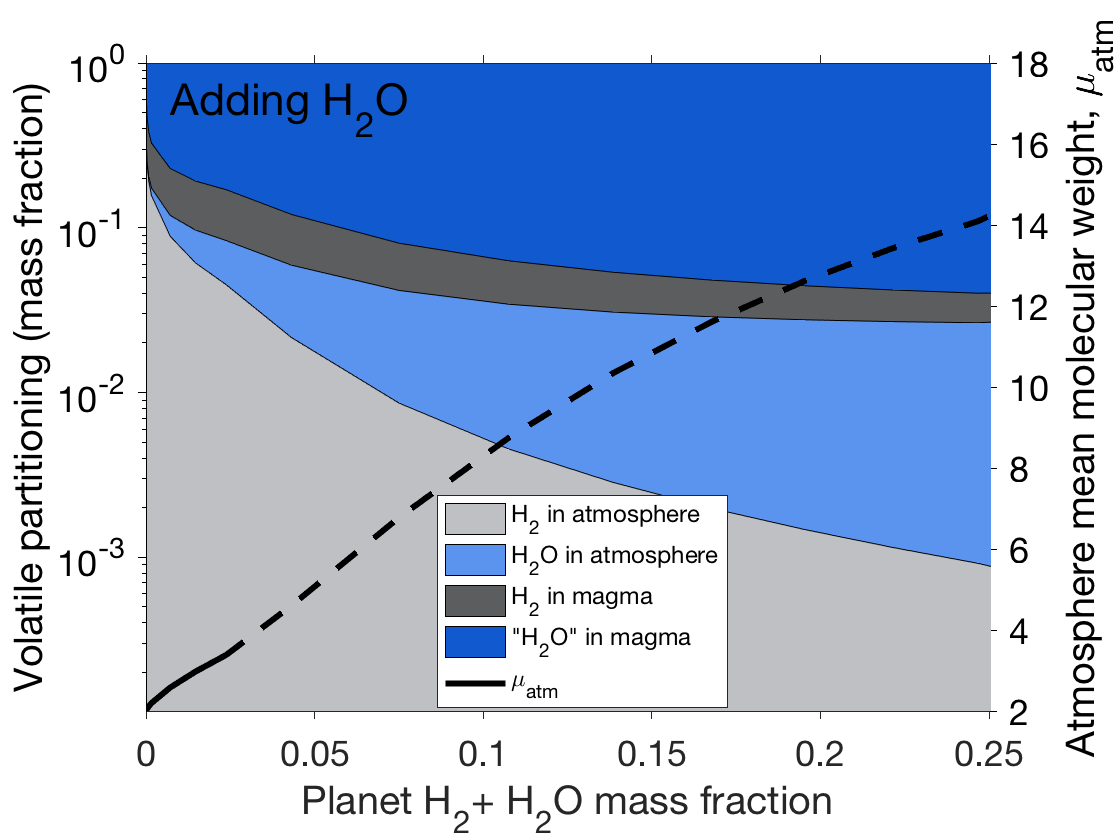}
                \centering
     \textbf{(d)}   \includegraphics[width=1.0\columnwidth]{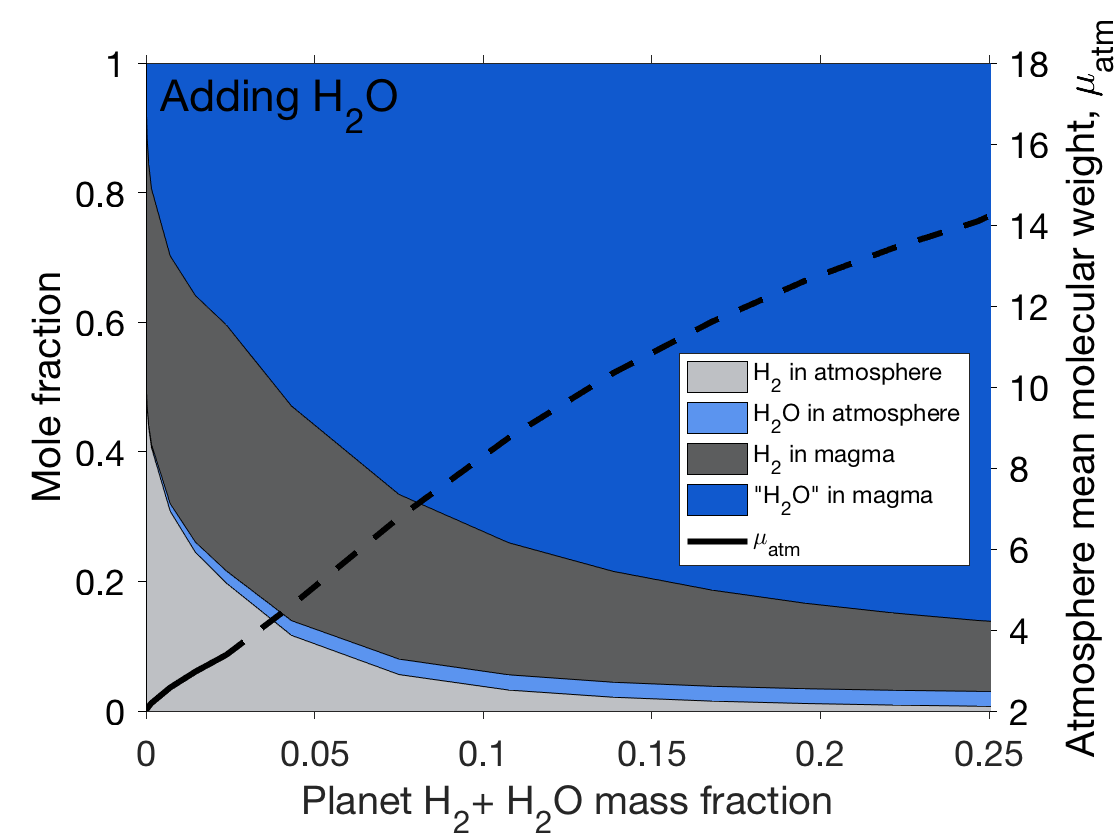}
\caption{How magma composition controls the fate of H: adding H$_2$O to a magma ocean containing Fe metal. Simplified model. Temperature at the  magma-atmosphere interface ($T_{mai}$) = 3000~K. For details, see \S3.2. \textbf{(a)} and \textbf{(b)} show the consequences of adding varying amounts of water to a planet with an initial dispersed-Fe-metal content of 50~wt\% in its magma ocean. (a) shows mass fractions and (b) shows mole fractions. 
\textbf{(c)} and \textbf{(d)}  show the consequences of adding varying amounts of water to a planet with an initial dispersed-Fe-metal content of 10~wt\% in its magma ocean. The point where enough H$_2$O has been added to oxidize all of the Fe is shown by the transition from the thick black solid line to the thick black dashed line.  (c) shows mass fractions and (d) shows mole fractions.} 
\label{fig:addh2oresults}
\end{figure*}

The results in Fig.~\ref{fig:addh2results} are for $T_{mai}$~=~3000~K. If decreasing $T_{mai}$ causes magma crystallization (this depends on planet $T_{eff}$, Fig.~\ref{fig:howmuchmagma}), then volatiles will be driven from the magma into the atmosphere -- a big effect. For simplicity, we assume mantles are fully molten here. For a fully molten magma and fixed atmospheric pressure, the net effects of cooling $T_{mai}$ from 3000~K to 2000~K are relatively small.

 Changing planet mass has little effect on atmospheric composition in our model (for fixed atmospheric pressure and fixed, high $T_{mai}$). This is because $R_{pl}$ is approximately proportional to $M_{pl}^{1/4}$ (we use $R_{pl}/R_\Earth  = (M_{pl}/M_{\Earth})^{0.27}$; \citealt{Valencia2006}). Thus gravity, $g \propto M_{pl}^{1/2}$. Since atmospheric mass = $P A_{mai} / g$ (where $A_{mai}$ is the area of the magma-atmosphere interface), atmospheric mass is almost independent of planet mass for fixed $P$.

\subsection{Adding Water to Fe-bearing Magma \textnormal{(Fig.~\ref{fig:addh2oresults}).}}

\noindent Could sub-Neptune atmospheres be the result of H$_2$ generation on the planet, with no need for H$_2$ delivery by nebula accretion? H$_2$ is generated when Fe metal reduces fluid H$_2$O \citep{ElkinsTantonSeager2008a,Genda2017,Haberle2019}. This H$_2$O is ultimately derived from solids, for example comets or hydrated asteroids.

Earth's ocean could be destroyed 400 times over by reaction with Earth's Fe-metal core  \citep{LangeAhrens1984}. Fortunately for life, most of the reducing power of Earth's core is safely buried \citep{Hernlund2016}. But is this true for exoplanets? 

We do not know whether or not H$_2$ generation by Fe-H$_2$O reaction on sub-Neptunes is  important. In this subsection we set a new constraint on the atmosphere boost by this process, by showing how atmosphere mean molecular weight increases as H$_2$O is added to an Fe-bearing magma ocean. Here we build on previous stochiometric calculations \citep{ElkinsTantonSeager2008a}, by adding the requirement of thermodynamic self-consistency. To motivate this upper-bound calculation, we first speculate on some scenarios by which the Fe-H$_2$O reaction might occur. We do not attempt to calculate how closely these scenarios approach our thermodynamic upper bound. Because we find that our new thermodynamic constraint is restrictive enough that most sub-Neptune atmospheres cannot be explained by Fe-H$_2$O reaction, the kinetics of these scenarios do not matter for the purpose of determining whether or not most sub-Neptune atmospheres can be explained by Fe-H$_2$O reaction.

Routes by which Fe might encounter H$_2$O on a sub-Neptune include the following. (1)~If most of the Fe-metal mass is delivered in the form of pebbles or planetesimals, then Fe-metal-bearing pebbles or planetesimals will vaporize in the atmosphere (e.g.~\citealt{Brouwers2018}), permitting chemical equilibration with the atmosphere. (2)~Small embryo impacts disperse Fe-metal through the magma ocean, increasing the chance that Fe-metal and H$_2$O can react \citep{Deguen2014}. (3)~Giant impacts can yield iron fragments, whose re-accretion promotes H$_2$ generation \citep{Genda2017}. (4)~For sufficiently-energetic giant impacts the boundaries between Fe-metal core, silicate magma, and the atmosphere become blurry \citep{Stevenson1984}. This physical boundary-blurring may favor chemical equilibration. 

 \begin{figure*}[t]
\textbf{(a)} \includegraphics[width=1.00\columnwidth]{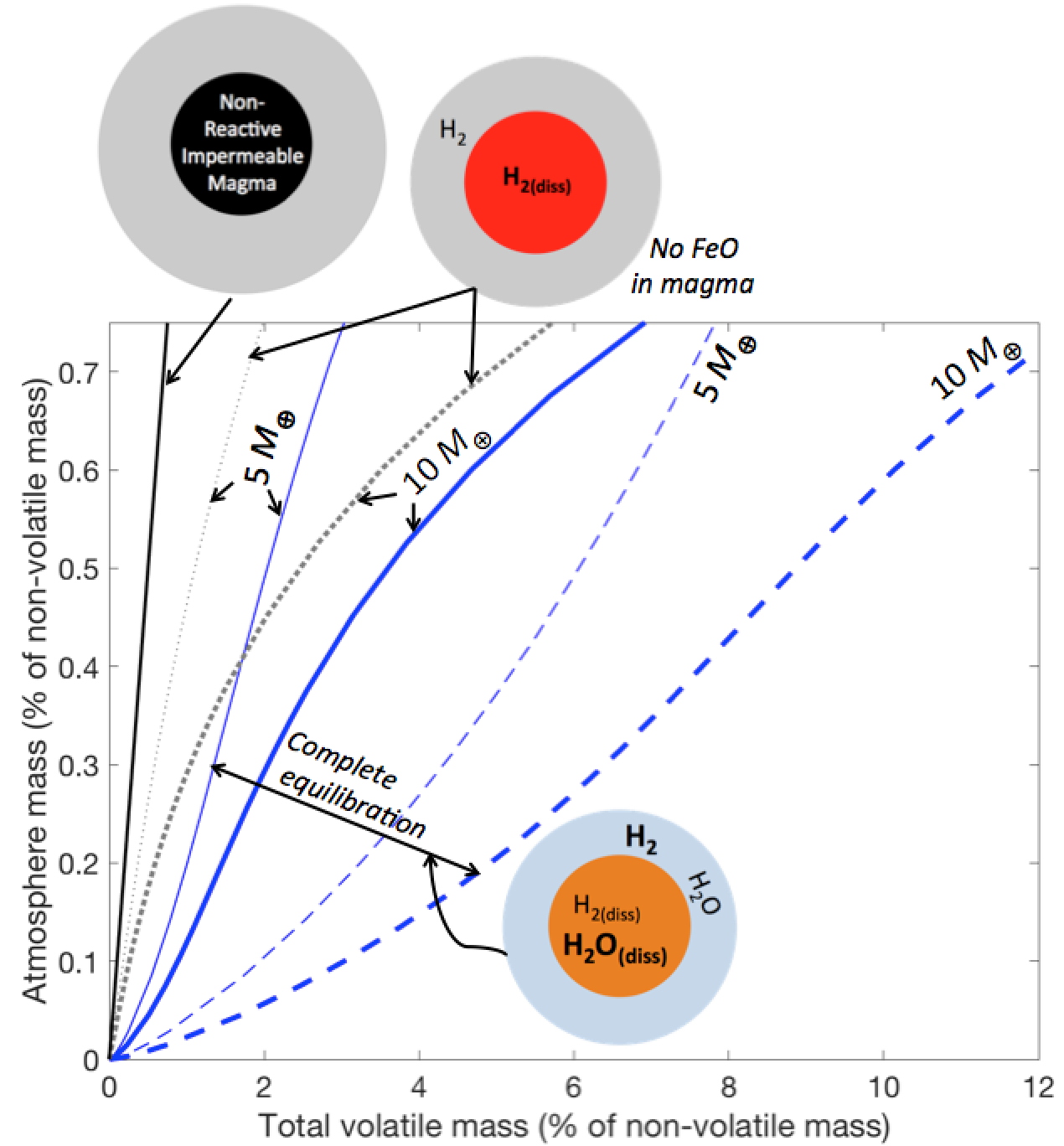}
\textbf{(b)} \includegraphics[width=1.05\columnwidth]{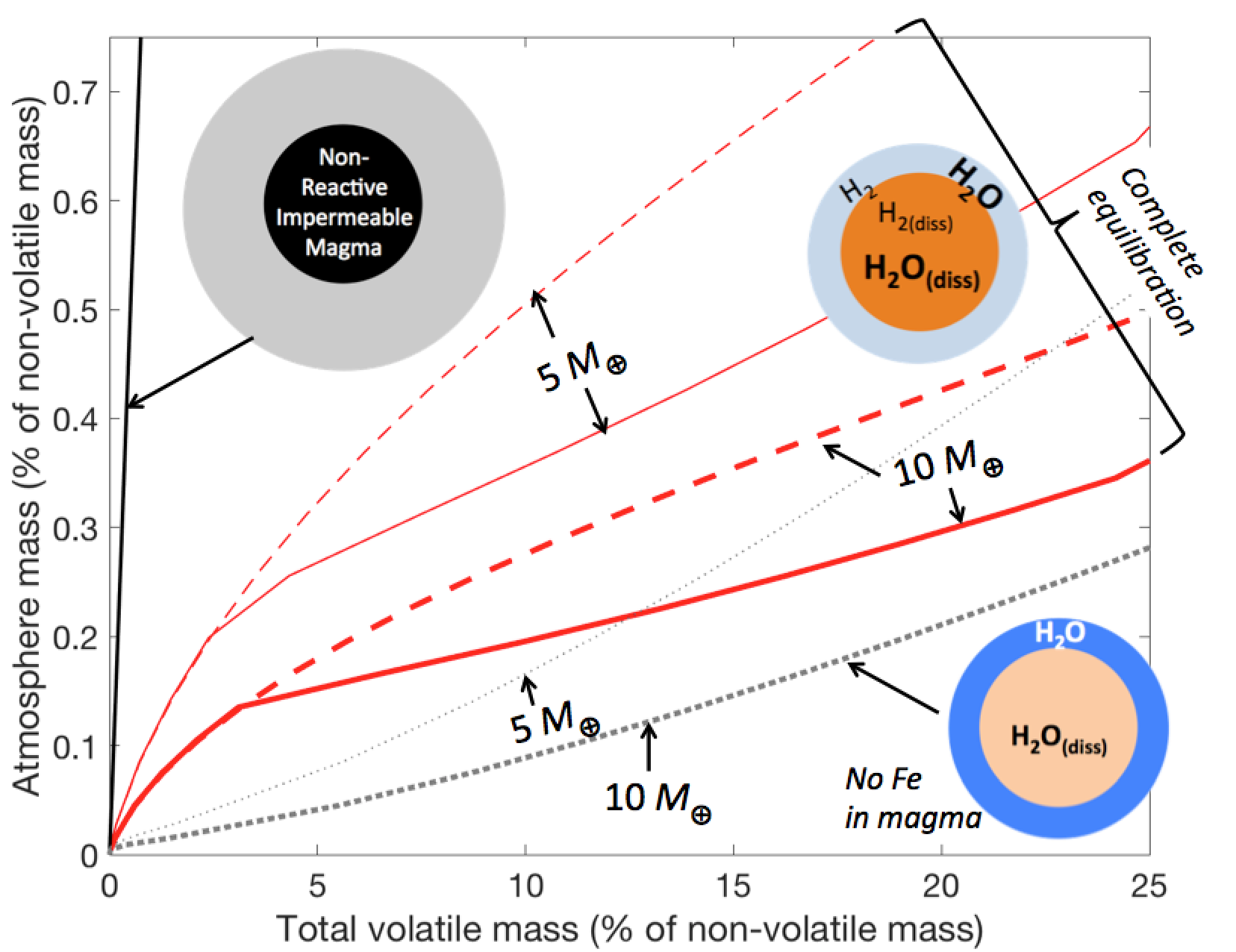} 
\caption{Atmosphere mass as a function of total volatile mass.  When the core is chemically non-reactive and impermeable, all volatiles are in the atmosphere (thick solid black line). \textbf{(a)} Blue lines show results for adding H$_2$ to magma with initial FeO content of 8~wt\% (solid lines) or 48.7~wt\% (dashed lines). As a set, the extrasolar silicate oxidation states reported by  \citet{Doyle2019} plot closer to the dashed lines than to the solid lines. Line thicknesses correspond to planet (silicate+metal) mass: thin for 5~$M_\earth$, thick for 10~$M_\earth$. Because H$_2$O is more soluble than H$_2$, the reaction FeO~+~H$_2$~$\rightarrow$~Fe~+~H$_2$O suppresses planet radius inflation - magmatic ``H$_2$O'' acts as a hydrogen sump. The gray lines correspond to a very reduced core (negligible FeO) - in this case, H is partitioned between dissolved H$_2$ and H$_2$ in the atmosphere (no H$_2$O). \textbf{(b)} Red lines show results for adding H$_2$O to a magma ocean with Fe content of 10~wt\% (solid lines) or 50~wt\% (dashed lines). Because H$_2$ is less soluble than H$_2$O, the reaction Fe~+~H$_2$O~$\rightarrow$~FeO~+~H$_2$ acts to boost the atmosphere relative to the reference case (which is shown by the gray dotted line). This reference case corresponds to an Fe-metal-free magma ocean. For details, see \S4.1.}  
\label{fig:5outcomes}
\end{figure*}

Results for adding water to a magma ocean initially containing 50~wt\% Fe metal (intermediate between the bulk-planet Fe~content of Earth and the bulk-planet Fe~content of Mercury) and with initially negligible FeO are shown in Fig.~\ref{fig:addh2oresults}a-b. Results for adding water to a magma ocean initially containing 10~wt\% Fe metal and initially negligible FeO are shown in Fig.~\ref{fig:addh2oresults}c-d. (10~wt\% Fe metal in the magma ocean might correspond to emulsification of the impactor after a giant impact onto proto-Earth.) H$_2$ is initially generated, but as the H$_2$O supply increases, the atmosphere molecular mass increases. Dissolved H$_2$O-in-magma is the dominant volatile reservoir. The results for different mantle Fe-metal contents are identical up to an atmosphere mass of $\sim$0.01~Earth~masses. Above that point, the two tracks diverge. This is because, for the 10~wt\% Fe-metal case (Fig.~\ref{fig:addh2oresults}c-d), all Fe-metal has been converted to FeO. In our model framework, no more H$_2$ can be generated. Adding more H$_2$O simply dilutes the H$_2$ generated by Fe~+~H$_2$O~$\rightarrow$~FeO~+~H$_2$. This dilution zone is shown by dashed lines in Figs. \ref{fig:addh2oresults}c-d, and corresponds to fast increase of atmosphere mean molecular weight. By constrast, for the 50~wt\% Fe-metal case, Fe~+~H$_2$O~$\rightarrow$~FeO~+~H$_2$ continues to release new H$_2$ into the atmosphere. This slows the rise to high atmosphere mean molecular weight.

Why is the composition of the first-produced atmosphere  independent of the Fe content for the H$_2$O-added-to-Fe-bearing-magma-ocean case, while the  composition depends on FeO content for the H$_2$-added-to-oxidized-magma case? This is because FeO exists as part of a liquid solution in the silicate magma (with MgO and SiO$_2$), whereas liquid Fe metal is a pure phase in our model \citep{Anderson1996}. As a result, the no-volatiles \emph{f}O$_2$ depends on the \emph{abundance} of FeO for an Fe-free magma, but for an FeO-free magma the no-volatiles \emph{f}O$_2$ depends only on the \emph{existence} of Fe.

\section{Analysis.}

\subsection{The Fraction of Volatiles That Are Stored in the Magma Is Variable, And This Decouples Radius From Composition \textnormal{(Fig.~\ref{fig:5outcomes})}.}

\noindent For H$_2$-dominated atmospheres, sub-Neptune radius is a proxy for atmospheric mass \citep{LopezFortney2014}.  However, radius is \emph{not} a proxy for total volatile mass if the atmosphere equilibrates with a massive magma ocean (Fig.~\ref{fig:5outcomes}). This is because of variable dissolution of H in the magma (as H$_2$ and as ``H$_2$O''). 

To see that dissolution matters, consider a planet where the volatile-free FeO content of the magma is the same as that of the Earth. Starting from a volatile-free world and adding H$_2$ until the atmosphere mass is 0.7~wt\%, we find that 3$\times$~more moles~of~H must be added (Fig.~\ref{fig:addh2results}b). This factor-of-3 difference is near the low end of our model predictions. Factor-of-50 (sic) enhancements are possible (Fig.~\ref{fig:5outcomes}).

To illustrate that atmospheric mass is not a good predictor of total volatile mass, we first make two restrictive assumptions. (\#1)~Nebular gas provides all volatiles. (\#2)~The range of  magma FeO contents (prior to any volatile addition) varies between planets, from [FeO]~=~0~wt\%, up to that of the Earth's [FeO], 8~wt\%. These two assumptions confine us to the triangular region between the thick black line and the solid blue lines in Fig.~\ref{fig:5outcomes}a. The uncertainty in total volatile mass (for a~given atmosphere mass) is a factor of $\gtrsim$10. Moreover, if we drop either of our two restrictive assumptions, then the uncertainty explodes. Mars and Vesta both have [FeO]~$\approx$~20~wt\%, and white dwarf data suggest exoplanet silicate [FeO] up to $\sim$50~wt\% \citep{Doyle2019}. If [FeO] can range from 0~wt\%~-~48.7~wt\%, then the uncertainty rises to a factor of $\sim$20 (the range between the thick black line and the dashed blue lines in Fig.~\ref{fig:5outcomes}a).  Alternatively, let us drop restrictive assumption~\#1. In this case, the volatiles could be predominantly solid-derived (Fig.~\ref{fig:5outcomes}b). For complete equilibration between magma and volatiles, the results are confined to the region between the gray dotted line and the dashed/solid lines in the lower right of Fig.~\ref{fig:5outcomes}b. But incomplete equilibration (or partial freeze-out) allows the planet to span the full range of parameters shown in Fig.~\ref{fig:5outcomes}b. The corresponding uncertainty is a factor of $\sim$100. 

Our analysis implies the following:

\begin{itemize}[leftmargin=*]

 \item Current theory (e.g. \citealt{Ginzburg2016,VenturiniHelled2017,Lee2018}) understates the amount of H that must be added to turn a rocky super-Earth-sized planet into a sub-Neptune by a factor of $\gtrsim$3, for a well-stirred deep magma ocean with an initial FeO content equal to that of Earth rocks (Fig.~\ref{fig:addh2results}).  %
 
 \item Escape processes used to explain conversion of sub-Neptunes into rocky super-Earth-sized planets \citep{Owen2019} must be more efficient by a factor of $>$3, for a well-stirred deep magma ocean with an initial (prior to H$_2$ addition) FeO content equal to that of Earth rocks. Magma stays equilibrated with the atmosphere as atmosphere is lost, so atmosphere losses will be mostly replaced by exsolution (Fig.~\ref{fig:5outcomes}) \citep{Schaefer2016}. 
 
\item If magma redox states are diverse \citep{Doyle2019}, then some close-in planets with radii in the super-Earth range will have a H$_2$ rich atmosphere. This is a novel prediction, for the following reason. A H$_2$-rich atmosphere must be thin ($<$0.05~wt\% of planet mass) for a $\ge$5~$M_\earth$ planet to remain in the super-Earth radius range. To explain such a thin atmosphere, fine-tuning of H inventory would be needed using the non-reactive impermeable magma approach. However, our model shows that such an atmosphere can correspond to 0.05-2~wt\% total volatile mass. With redox diversity, no fine-tuning of H inventory is needed to give a H$_2$ rich atmosphere for planets with radii in the super-Earth range. This prediction can be tested with ARIEL \citep{Tinetti2016} and perhaps JWST (at HR~858; \citealt{Vanderburg2019}; or at GJ~9827c; \citealt{Rice2019}). 

\end{itemize}

\begin{figure}
\includegraphics[width=1.025\columnwidth,clip=true,trim={3mm 0mm 0mm 0mm}]{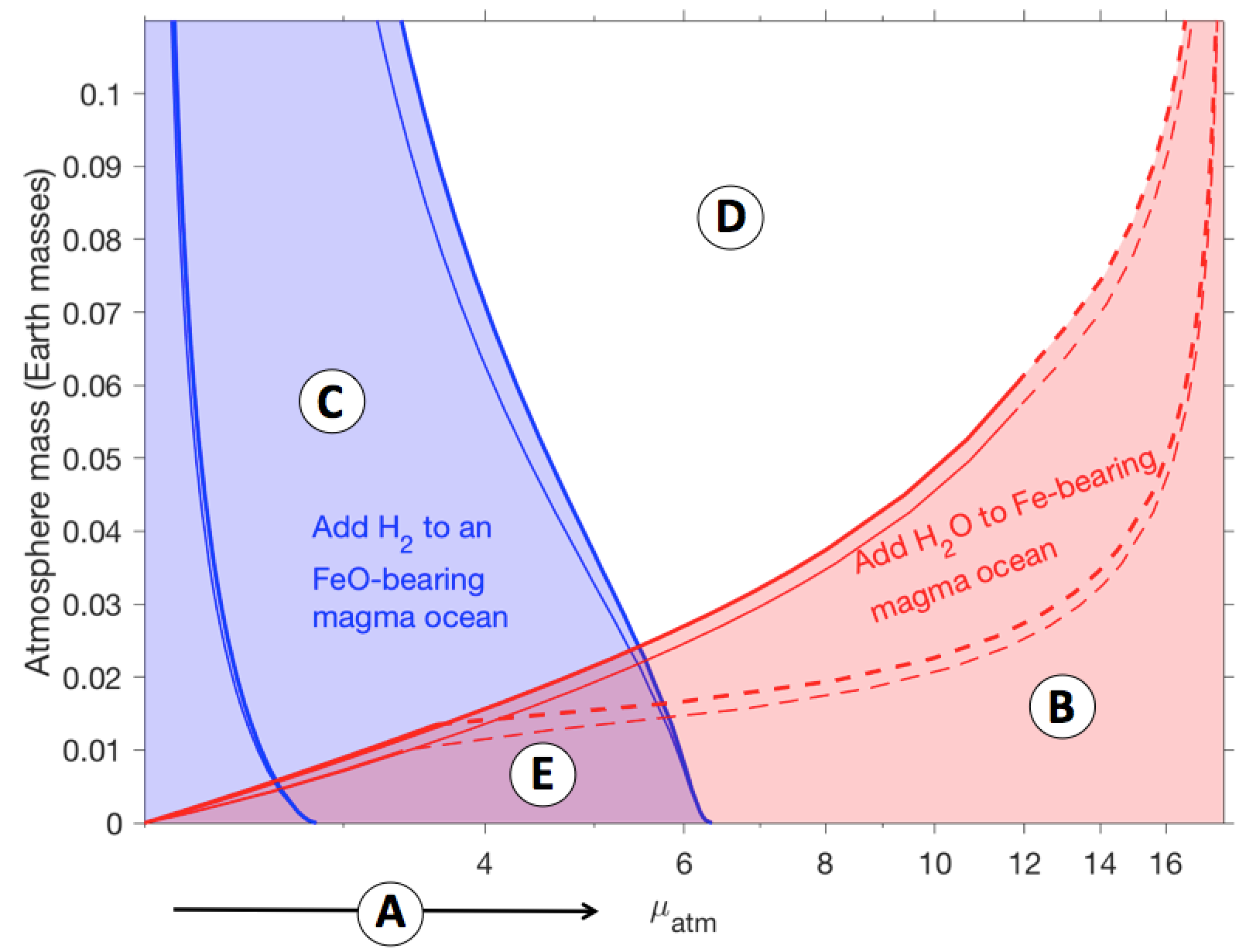}
\caption{Effect of magma-atmosphere reactions on sub-Neptune atmosphere mass and atmosphere mean molecular weight (simplified model, at magma-atmosphere interface temperature $T_{mai}$ = 3000~K, assuming a well-mixed H$_2$/H$_2$ atmosphere). Thin lines: $M_{pl}$~=~5~$M_\Earth$, thick lines: $M_{pl}$~=~10~$M_\Earth$. Arrow \mycirc{\textsf{A}} corresponds to the increase in $\mu_{atm}$ from atmosphere reactions with an oxidized core, or supply of solid-derived volatiles. Red zone \mycirc{\textsf{B}}  corresponds to adding H$_2$O-rich solid-derived volatiles to an Fe-bearing magma ocean. Blue zone \mycirc{\textsf{C}} corresponds to adding nebular gas to an FeO-bearing magma ocean. Adding nebular gas to a magma ocean with the FeO contents inferred for extrasolar silicates from white dwarf spectra by \citep{Doyle2019} would yield atmospheres in the right-hand half of the blue zone. White zone \mycirc{\textsf{D}} is water-buffered. Purple zone \mycirc{\textsf{E}} corresponds to ambiguous atmospheric origins. The leftmost pair of blue lines corresponds to an Earth-like volatile-free mantle FeO content. The rightmost pair of blue lines corresponds to a volatile-free mantle FeO content of 48.7 wt\%.  The leftmost pair of red lines correspond to a magma ocean that has 50 wt\% Fe~metal in the volatile-free limit. The rightmost pair of red lines correspond to a magma ocean that has 10 wt\% Fe~metal in the volatile-free limit. The point where enough H$_2$O has been added to oxidize all of the Fe corresponds to the transition from the red solid lines to the red dashed lines. We did not model Fe$^{3+}$, but if we had, then both the colored zones would slightly expand. For details, see \S4.2.}
\label{fig:muboost}
\end{figure}

\subsection{Measurements of Atmosphere Mass and Mean Molecular Weight Can Probe Atmosphere Origins \textnormal{(Fig.~\ref{fig:muboost}).}}

\noindent Sub-Neptune formation is not well understood. We do not know how rocky cores grow -- giant impacts, or pebble/planetesimal accretion (e.g., \citealt{ChatterjeeTan2014, Levison2015})? We do not know where the growth happened -- formation in-situ, or planet migration (e.g., \citealt{ChiangLaughlin2013, Mordasini2018})? We do not know where the volatiles came from -- nebula gas, or solid derived volatiles? 
 
These hypotheses make distinct predictions for the mean~molecular~weight of the atmosphere ($\mu_{atm}$) and the atmosphere mass / planet mass ratio ($f_{atm}$). Both can potentially be constrained by observations (e.g. \citealt{BennekeSeager2012,Fortney2013,Benneke2019a}). Within our model, for $T_{eff}$~$>$~400~K sub-Neptunes, we make the following connections (Fig.~\ref{fig:muboost}):

\begin{itemize}[leftmargin=0em]

\item[]  \mycirc{\textsf{A}} \emph{High H$_2$O/H$_2$ ratio (much greater than solar) $\rightarrow$ planetesimal-or-planet migration, or Si dissolution into Fe-metal core}. A~high H$_2$O/H$_2$ ratio in the atmosphere (much greater than solar) indicates the magma is oxidized.  Oxidized magma can be produced either by net oxidization of the planet's materials, or by rearrangement of oxidizing power inside the planet, with a net transfer of reductants to the Fe-metal core leaving oxidants in the silicate mantle (Appendix~B). Net-oxidation of a planet's materials involves a contribution to the planet's building blocks from beyond the water-ice snowline -- either planet migration, or planetesimal migration. This is because one easy way to oxidize a planet's building blocks is the reaction Fe$_{\mathrm{(s)}}$+ H$_2$O$_{\mathrm{(l)}}$ $\rightarrow$ FeO$_{\mathrm{(s)}}$ + H$_{\mathrm{2(g)}}$, followed by escape of H$_2$ to space (Appendix B). Thus magma oxidation should (for an ensemble of planets) increase with semimajor axis (\citealt{Rubie2011,Rubie2015}; Appendix~B). This tendency is seen in the Solar System: Mercury's mantle has $\le$0.04~wt\%~Fe$^{(2+)}$O \citep{Nittler2017}, Earth's has $\sim$8~wt\%~Fe$^{(2+)}$O, Mars' has 18~wt\%~Fe$^{(2+)}$O, and Vesta's mantle has $\sim$20~wt\%~Fe$^{(2+)}$O. If (near-)resonant chains of planets record more migration than planetary systems that lack such chains, then systems with (near-)resonant chains of planets should have more-oxidized magma oceans.

The Si-dissolution mechanism predicts that planets that formed hotter (i.e. bigger planets) will tend to have magmas that are more oxidized \citep{Fischer2017} (Appendix~B).

\item[]  \mycirc{\textsf{B}} \emph{ $\mu_{atm}$ $>$ 7 $\rightarrow$ large-object migration}. Sub-Neptune atmospheres cannot reach $\mu_{atm}$~$>$7 by reactions between magma and nebula gas (Fig.~\ref{fig:muboost}). Instead, gas from solid-derived volatiles (H$_2$O, C-species, e.t.c.) is needed. Because the icelines for these solid-derived volatiles are at orbital period $>$~10$^2$ days, $\mu_{atm}$~$>$~7 implies inward migration of objects that are large enough to resist dry-out during migration \citep{Bitsch2019}. Atmospheres formed by reacting liquid Fe-metal with solid-derived volatiles plot in the red region in Fig.~\ref{fig:muboost}. Atmospheres derived directly from solid-derived volatiles can also plot in this region. The two options might be distinguished by probing for C species. C species are extremely Fe-loving \citep{DasguptaGrawal2019}, so if atmospheres interact with liquid Fe-metal, then C species should be absent. Liquid iron scrubs C from the atmosphere.

\item[]  \mycirc{\textsf{C}} \emph{$\mu_{atm}$ $<$ 7, plus atmosphere mass $>$0.03 $M_\Earth$  $\rightarrow$ nebula accretion}. Points in Fig.~\ref{fig:muboost} that are within the blue zone, above the red line, can only be formed by nebular accretion (not generation of H$_2$ by Fe-H$_2$O reactions). Within this region, upper limits on $\mu_{atm}$ (from the amplitude of features in transmission spectra) place upper limits on the [FeO] content of magma, and thus on the extent of atmosphere-magma interaction. This is because planets with a higher volatile-free magma FeO content, or that have a greater degree of atmosphere-magma equilibration, plot further to the right within the blue zone on Fig.~\ref{fig:muboost}.

\item[]  \mycirc{\textsf{D}} \emph{Water-buffered worlds.} Atmospheres that plot in the white zone in Fig.~\ref{fig:muboost} cannot be explained by gas release by reaction of arriving material with the magma. These worlds likely gathered a major contribution from H$_2$O.

\item[]  \mycirc{\textsf{E}} \emph{Zone of overlap = ambiguous origins.} Thin atmospheres with 2 $<$ $\mu_{atm}$ $<$ 7 can be explained either by endogenic generation of H$_2$ or by nebular accretion.  Thin atmospheres can also be explained by high-molecular-weight volcanic outgassing, as on Earth, Venus, and Mars.

\end{itemize}

\noindent Fig.~\ref{fig:muboost}~relates~the mass, and the mean~molecular~weight, of sub-Neptune atmospheres to two key parameters. The first parameter is the magma redox state. The second parameter is the number of H atoms contained within the sub-Neptune.  Both parameters can constrain models of  planet formation. 

In future, further constraints might come from D/H data, where D/H values that are elevated (relative to the D/H of the host star) point toward a greater contribution from solid-derived volatiles \citep{Morley2019}.

\subsection{Endogenic H$_2$ cannot explain most orbital period~$<$~100~d sub-Neptunes.}

Can the radii of sub-Neptunes be explained by H$_2$ generated on the planet (endogenic H$_2$)? This is (just) possible if every Fe atom gives up all three redox-exchangeable electrons to form H$_2$  from H$_2$O \citep{ElkinsTantonSeager2008b,RogersSeager2010,Rogers2011}. In that case, up to 3.6~wt\%~H$_2$ will be generated. 3.6~wt\%~H$_2$ is sufficient to account for most sub-Neptune radii, if all H$_2$ resides in the atmosphere. However, most of the H$_2$ will dissolve into the magma (Fig.~\ref{fig:5outcomes}). Moreover, our model shows that adding H$_2$O will lead to a high-molecular-mass atmosphere ($\mu$~$>$~8) for atmosphere mass $>$0.8~wt\% of planet mass (Fig.~\ref{fig:muboost}). Therefore, endogenic H$_2$ cannot explain most orbital~period~$<$~100~d sub-Neptunes. Nebula gas is needed.

\subsection{Trends over Gyr: Puff-Up versus Late Ingassing}

Existing models predict that sub-Neptunes will shrink with age as they cool and lose mass  (e.g., \citealt{LopezFortney2014,Vazan2018b}). Magma-atmosphere equilibration makes different predictions. These predictions can be tested by comparing radius data for planets that differ in age. The needed data are now becoming available (e.g.,~\citep{SilvaAguirre2015,Johnson2017, Mann2017, Newton2019,David2019}). 

\emph{Puff-Up:} 
As planets cool and/or lose H$_2$ to space, the magma ocean can freeze (Fig.~\ref{fig:howmuchmagma}). The magma-atmosphere interface is not the first layer to freeze. Rather, the freezing starts at great depth (e.g.,~\citealt{Bower2019}). When magma starts to freeze, volatiles are excluded from the solid and are strongly enriched in the residual melt. The volatile-enriched melt forms bubbles, which burst at the magma-atmosphere interface. Magma ocean freezing will squeegee volatiles into the atmosphere \citep{ElkinsTanton2011}.

Volatiles are pistoned into the atmosphere (leading to bigger radii than predicted by existing models) on the freeze-out timescale (billions of years; e.g.~\citealt{Vazan2018b}). This puff-up effect depends on semimajor axis (because shorter-period worlds will never freeze) and on atmospheric thickness (because thickly-blanketed worlds will never freeze). Puff-up predicts a statistical excess of sub-Neptunes in the diagonal band between 2 Earth masses of magma and 0.1 Earth masses of magma on Fig.~\ref{fig:howmuchmagma}. This prediction might be tested with TESS mission extensions, or  PLATO \citep{ESA2013}.

\emph{Late Ingassing:} Consider a sub-Neptune with a H$_2$ atmosphere that is initially \emph{out}-of-equilibrium with an FeO-rich magma ocean. This might occur if the magma is initially stratified (no convection), and the H$_2$ accretes only after the magma+metal core is assembled. In this scenario, the deeper magma will stay volatile-free until the uppermost part of the magma ocean cools enough for the magma to convect (timescale $\gtrsim$\emph{O}(10$^{9}$) yr;  \citealt{Vazan2018b}). Once convection starts, the planet will shrink. This is because (1) H$_2$ will dissolve into the magma and (2)~as~the atmosphere is progressively oxidized, it becomes more soluble. Late ingassing predicts a shrinkage excess relative to existing models.

H$_2$O generation by the redox reactions could slow the cooling of planets with nebular-derived atmospheres. This is because H$_2$O is a major source of radiative opacity.

\section{Discussion.}

\subsection{Approximations and Limitations}

\subsubsection{Material properties} Our material properties are mostly extrapolated from $T$~$<$~2000~K lab data. At higher temperatures, they could be in error (e.g. \citealt{FegleySchaefer2014, Fegley2016,SossiFegley2018}). For example, we~neglect $T$-dependence of H$_2$O solubility. This dependence is weak (for Si-poor magmas) at $T$~$<$~2000~K \citep{FegleySchaefer2014}.   More lab experiment and/or numerical-experiment data (e.g. \citealt{SoubiranMilitzer2015})  for material properties (especially solubility of volatiles in silicate) at the $T$ and $P$ of magma-atmosphere interfaces on sub-Neptunes are critically needed. 

Our H$_2$O solubility model is for dissolution in basaltic magma because high-pressure data for peridotite magma is not available. H$_2$O solubility in magma increases dramatically with total pressure, and water and (peridotitic) magma are fully miscible above 3-6 GPa \citep{ShenKeppler1997, Ni2017}. Non-linear H$_2$O solubility at much lower pressures interpolates between the solubility at low water content (for which H$_2$O is dissolved mainly as OH$^-$ and $\sqrt P$ solubility is a good match to data; \citealt{AbeMatsui1986,Pohlmann2004, Karki2010}), and the solubility at high water content (for which H$_2$O is increasingly dissolved as molecular H$_2$O;  \citealt{Stolper1982}).  The \emph{f}O$_2$-dependence of H$_2$O solubility is weak for Fe$^{3+}$-absent conditions  \citep{BolfanCasanova2002}, and we ignore it.
   
\subsubsection{Composition} If we had tracked carbon instead of just hydrogen \citep{Bitsch2019}, then adding solid-derived volatiles would have produced a quicker increase in $\mu_{atm}$. 

Our toy model of magma ocean mass uses melting curves reported for a ``chondritic mantle'' composition with minimal volatiles \citep{Andrault2011}. This composition can be thought of as average Solar System rock \citep{Unterborn2017,PutirkaRarick2019}. Including the effect of volatiles on the melting curve would favor melting, and so increase magma ocean mass \citep{Katz2003}. On the other hand, reported volatile-free pyrolite melting curves are at higher-$T$ than for chondritic mantle \citep{Andrault2017}; switching from a chondritic-mantle melting curve to a pyrolite melting curve would reduce magma ocean mass.

We~neglect Fe$^{3+}$ \citep{KressCarmichael1991, Frost2004, Zhang2017}. Including Fe$^{3+}$ would modestly increase the compositional effect of magma-atmosphere interaction. Therefore, our omission of Fe$^{3+}$ is conservative. 

\subsubsection{Thermodynamic treatment} 

We~neglect H storage by dissolution into liquid Fe-metal \citep{Stevenson1977,Wu2018,Clesi2018}. If this reservoir is large, then that would strengthen our conclusion that only a fraction of the H supplied from the nebula stays in the atmosphere.

We omit helium. Helium may slightly decrease H$_2$ solubility in melt, but because of helium's small mole fraction in nebular gas, this is unlikely to be a big effect. 

For the atmosphere, we use the Lewis-Randall approximation. In other words, we neglect the fact that the fugacity coefficient of a gas in a mixture is different than that of the pure gas due to the molecular interactions in the gas mixture. 

For our Fe-FeO buffer calculations, in effect we assume the partial molal volume of FeO in magma is independent of the magma Si content. Our assumed partial molal volumes for FeO are from measurements on melts that have more Si than for plausible sub-Neptune magma ocean compositions \citep{Armstrong2019}. This is probably not a big effect.

Thermal dissociation of H$_2$ to atomic H is minor for our purposes. At 3000~K and 10$^8$~Pa, the H/H$_2$ molar ratio is just $\sim$0.5\%, dropping further at higher pressures. 


\subsection{How Long For Magma-Atmosphere Equilibration?}   Our calculations assume magma-atmosphere equilibration happens during planet formation. We do not know whether or not this assumption is correct. This assumption is reasonable if magma oceans grow by giant impacts (e.g. \citealt{InamdarSchlichting2015}), and each giant impact energetically stirs the growing planet. The equilibrium assumption is also plausible if magma oceans grow by condensation and rain-out of solids that were vaporized upon accretion (e.g., \citealt{Brouwers2018,Bodenheimer2018}). However, this planetesimal-or-pebble-accretion scenario might give rise to a stably-stratified magma ocean, with hot H-rich layers overlying deeper rock layers that are H-poor and cooler because they equilibrated with lower-pressure atmospheres earlier in planet growth history. Such an onion-shell structure would have less H$_2$ dissolved in the magma, and (for a nebula volatile source) less H$_2$O in the atmosphere, than for our default model. Finally, the magma ocean might form volatile-poor if silicate cores reach full size in an environment that has a low base-of-atmosphere pressure (e.g., \citealt{Ormel2015}). In this last case, convection is needed for non-negligible equilibration \citep{Pahlevan2019}. Dissolution of volatiles into magma decreases the density of magma \citep{OchsLange1999}, an effect whose sign is to stabilize magma layers near the top of the magma ocean (disfavoring convection). Convection will cease if the stabilizing buoyancy from gradients in volatile abundance with depth exceeds the destabilizing buoyancy from planet cooling. Even if convection continues, the volatile-enriched boundary layer is thinner than the thermal boundary layer (because the diffusivities of H and OH are less than the thermal diffusivity). This slows regassing.

There is another way to suppress magma convection. At~sufficiently high temperature, iron, melt and volatiles become fully miscible. In this limit, magma and atmosphere are indistinguishable -- a single phase. For example, water and (peridotitic) magma are fully miscible above 3-6 GPa \citep{Ni2017}.  In the~context of planet formation, full miscibility at high $T$ suggests a zone of intermediate density \citep{HelledStevenson2017,Bodenheimer2018,BrouwersOrmel2019}. Such a rock-volatile fuzzy zone has recently been discovered on Jupiter \citep{Wahl2017}. As fuzzy zones cool, the temperature gradient favors convection, but the compositional gradient inhibits convection. For H$_2$/magma fuzzy zones, it is not clear whether or not convection  can continue (a review of the relevant fluid mechanics problem is given in \citealt{Garaud2018}; parameterizations of the consequences of sluggish or stalled convection for planet thermal evolution include that of \citealt{LeconteChabrier2012}; and \citealt{FrenchNettelmann2019} calculate the crucial Prandtl number, albeit for H$_2$O). If convection stalls and does not restart, then magma-atmosphere equilibration will effectively stop.

\begin{figure}
\centering
\includegraphics[width=1.01\columnwidth,trim={0mm 0mm 0mm 0mm}]{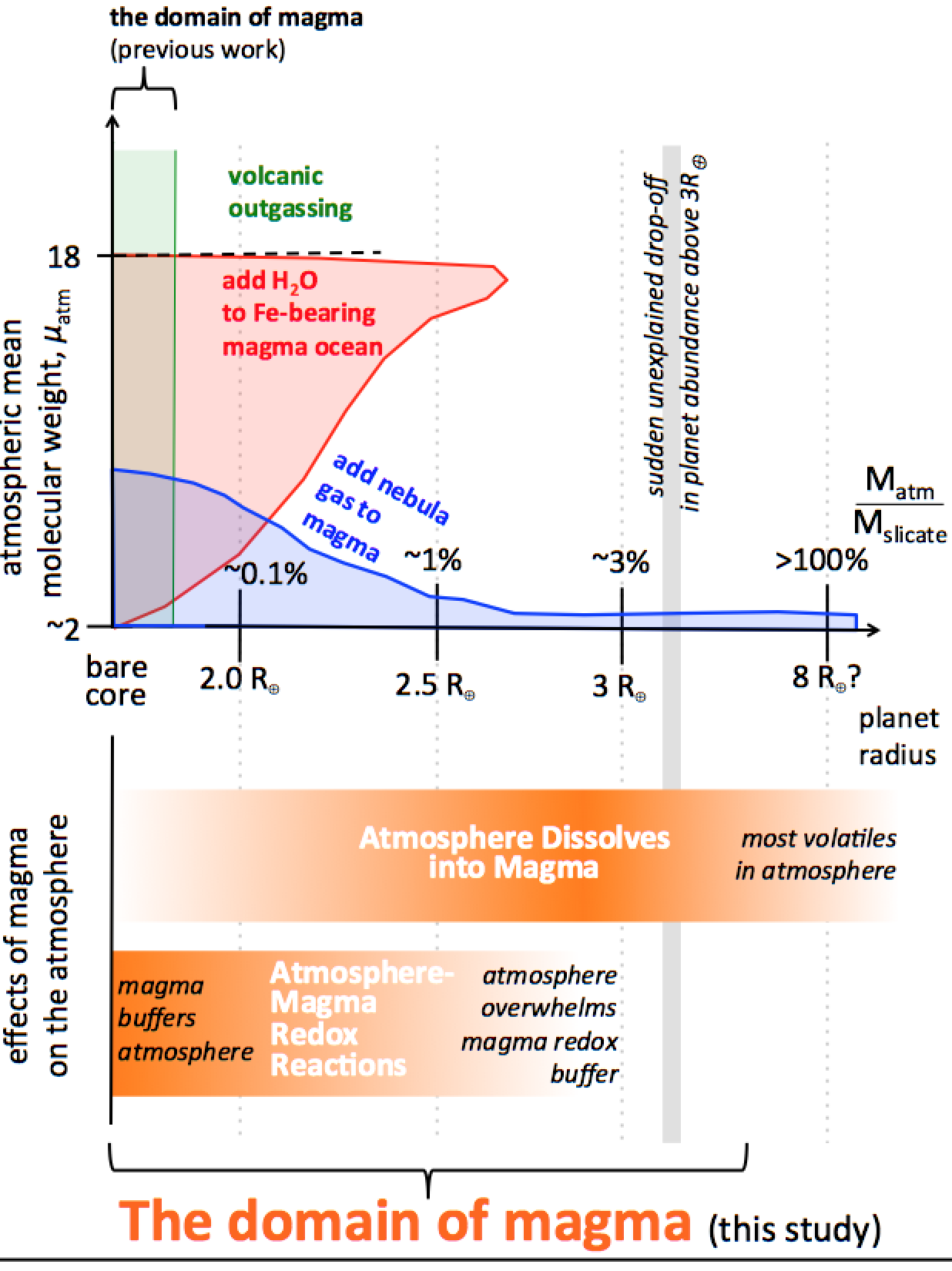}
\caption{Conclusions in context. We found that both magma-atmosphere redox reactions and atmosphere dissolution into magma can be important for setting atmospheric composition. For very massive atmospheres, the redox buffer is overwhelmed, but dissolution of atmosphere in magma is very important \citep{Kite2019}. At sufficiently high planet radius, the atmosphere greatly outweighs the silicates and the atmosphere is little-affected by the silicates. For atmosphere mass $\ll$0.1 wt\% of planet mass, volcanic outgassing (not considered in this paper) may explain atmospheres. The bottom line is that our work expands the range of atmosphere compositions and atmosphere masses that can be explained by atmosphere interactions with magma.}  
\label{fig:beforeafter}
\end{figure}

\section{Conclusions.}

We draw five main conclusions from within the framework of our simplified model. 

\begin{enumerate}[leftmargin=*]
\item Magma matters. Sub-Neptune atmosphere composition and mass can be greatly affected by atmosphere-magma interaction (Figs. \ref{fig:cartoon}-\ref{fig:addh2oresults}). For worlds that are mostly magma by mass, atmosphere H$_2$O/H$_2$ records both volatile delivery (exogenic water) and also volatile-magma interactions (which may produce endogenic water). The need to consider endogenic water complicates interpretation of atmosphere metallicity measurements (e.g., \citealt{Mordasini2016}). Atmosphere metallicity measurements have already been made for Neptune-sized exoplanets (e.g., \citealt{Fraine2014,CrossfieldKreidberg2017, Wakeford2017,Morley2017,Turrini2018,Benneke2019a,Benneke2019b}), and will soon be extended to sub-Neptunes. Interpretation of atmospheric metallicity on sub-Neptunes will not be a simple extension of theory developed for gaseous planets.

\item  A lot of H goes into the magma \citep{ChachanStevenson2018} (Figs.~\ref{fig:addh2results}-\ref{fig:5outcomes}). Boosting the radius of a magma-cored sub-Neptune requires more H$_2$ than is usually assumed. For example, if sub-Neptune silicates have a volatile-free magma FeO content similar to that inferred for extrasolar silicates on the basis of white dwarf spectra \citep{Doyle2019}, then, for full magma-atmosphere equilibration, $>$4$\times$ more H$_2$ must be added to explain a given sub-Neptune's radius.

\item  Turning a sub-Neptune into a rocky super-Earth-sized planet requires more H loss than is usually assumed. This is because H loss is compensated by exsolution from the magma. Bubbling is a negative feedback on atmospheric loss. The corresponding increase in the demand for H loss (to explain a given radius change) pushes hypothesized H loss mechanisms closer to their energetic upper limits. This extra demand may stress-test H loss models such as core-powered mass loss \citep{Ginzburg2018}, impact erosion \citep{InamdarSchlichting2016, ZahnleCatling2017, BierstekerSchlichting2019}, and photoevaporation \citep{OwenWu2017}.

\item Atmosphere mass and atmosphere mean molecular weight can be used as a proxy for atmosphere origin (Fig.~\ref{fig:muboost}). Atmosphere H$_2$O/H$_2$ ratio is proportional to magma FeO content. 

\item If the magma and the atmosphere equilibrate, as is assumed by our model, then there are consequences we can test \citep{Kite2019}. One example is a statistical excess of sub-Neptunes with $\sim$10$^{25}$~kg magma oceans (\S4.4). 
\end{enumerate}

Our model is limited by the lack of material properties data (e.g., solubilities) from lab and/or numerical experiments for the conditions at the sub-Neptune magma-atmosphere interface ($T$~$>$~2000~K, $P$~=~1-10~GPa). The effects that we have uncovered are big, so such lab and/or numerical experiments are strongly motivated if we are to understand what lies at the heart of sub-Neptune composition and evolution.  

\acknowledgments \noindent \emph{Acknowledgements.} 
We thank an anonymous reviewer for useful comments. We thank Paul Asimow, Andy Campbell, Dan Fabrycky, Greg Gilbert, Marc Hirschmann, Nadia Marounina, Miki Nakajima, Megan Newcombe, Leslie Rogers, Dave Stevenson, and Sarah Stewart, for discussions. We thank Kaveh Pahlevan for feedback on a draft. This work was supported by the U.S. taxpayer, primarily via grant NNX16AB44G (NASA), and secondarily via grant NNX17AC02G (NASA) and grant AST-1517541 (NSF). E.B.F.~ acknowledges support from the Center for Exoplanets and Habitable Worlds, which is supported by The Pennsylvania State University, the Eberly College of Science, and the Pennsylvania Space Grant Consortium. 
\vspace{0.1in}

\noindent \emph{Author contributions:} E.S.K. conceived research. E.S.K. and B.F. designed research. E.S.K. carried out the work, with contributions from B.F. (fugacity coefficients) and from L.S. (Fe-FeO buffer values); and E.S.K. wrote the paper, with a contribution (Appendix~C) from B.F. All authors contributed to editing the paper.

\vspace{0.1in}
\noindent \emph{Code availability.} Everything used to make this paper can be obtained for unrestricted further use by emailing the lead author. 
\vfill\null


%
 
\section*{Appendix~A: \\ Magma on sub-Neptunes}
\noindent Rock with the composition of average Earth rock starts to melt at 1350~K and is mostly molten by 1750~K \citep{Katz2003}. Sustaining such high temperatures (for orbital period~$>$~3~days) requires an atmospheric blanket \citep{IkomaGenda2006}. To find the atmospheric-blanket thickness, we first note that the most important constituent of the deep (convecting) atmosphere of sub-Neptunes is molecular hydrogen, H$_2$. We use an adiabatic index for H$_2$, $\gamma$, of 4/3, which is appropriate for our pressure ($P$) and temperature ($T$) of interest \citep{Saumon1995}. Then for an ideal gas we have
 
 \begin{equation}
\left(\frac{T_{mai}}{T_{RCB}}\right) = \left(\frac{P_{atm}}{P_{RCB}}\right)^{\frac{\gamma - 1}{\gamma}} \approx \left(\frac{P_{atm}}{P_{RCB}}\right)^{0.25}
\label{eqn:adiabat}
 \end{equation}

\noindent where ``\emph{mai}'' denotes the magma-atmosphere interface, and $_{RCB}$ denotes the boundary between the radiative outer atmosphere, and the convecting (adiabatic) deep atmosphere (Fig.~\ref{fig:pieslice}). Eqn.~\ref{eqn:adiabat}, with $P_{RCB}$~=~10~bars, gives us the colored curves in Fig.~\ref{fig:howmuchmagma}. We assume that the planet's internal luminosity is small compared to insolation, but is still large enough that the radiative zone is small. Thus $T_{RCB}$ $\approx$ $T_{eff}$ (effective temperature) $\approx$ $T_{eq}$ (equilibrium temperature). 

To obtain the mass of the magma (the dashed lines in Fig.~\ref{fig:howmuchmagma}), we must consider another effect: under pressure, magma will freeze. To track pressure-freezing, we used Fig.~5 in \citet{Andrault2011} to get the $T$($P$) for the 100\%-melt curve (the liquidus), the 0\%-melt curve (the solidus), and magma adiabats. We interpolated linearly in tempeature for intermediate melt fractions. (The liquidus and solidus both curve significantly at high pressure and are almost parallel to the adiabat around 10$^2$~GPa; \citealt{Bower2018,MiyazakiKorenaga2019}). This gave us a relationship between $P$ and melt fraction for a range of $T_{mai}$. (We assume that the  temperature difference between the top of the magma and the lower layers of the atmosphere is small.) To map $P$ onto depth within the silicate interior of sub-Neptunes, we assumed the relationship between pressure and density for the Earth's silicate mantle (from \citealt{DziewonskiAnderson1981}; magma densities are up to 15\% lower, \citealt{Bower2019}). We integrated the pressure downward from the top of the magma, assuming the distance from the center of the planet to the magma-atmosphere interface $R_{mai}$ is given by 

\begin{equation}
\left(\frac{R_{mai}}{R_\earth}\right)  = \left(\frac{M_{pl}}{M_\Earth}\right)^{0.27}
\label{eqn:rplmpl}
\end{equation}

\noindent \citep{Valencia2006}, where $M_{pl}$ is planet mass.

Above $T_{mai}$ = 3000K, a mostly or fully molten mantle is a reasonable approximation for sub-Neptune mass $\le$4~$M_\earth$ (Fig.~\ref{fig:howmuchmagma}). This result is specific to the melting curves reported by \citet{Andrault2011}. 

The atmosphere thermal blanket effect is much more important than pressure freezing from the weight of the atmosphere. As a result, the dashed lines in Fig.~\ref{fig:howmuchmagma} are nearly parallel to the  $T_{mai}$ contours.

Magma ocean mass depends strongly on $T_{mai}$, but only weakly on planet mass. For thin magma shells, $M_{magma}$~$\propto$~$A/g$, $M_{magma}$~$\propto$~$M_{pl}^{0.08}$ from Eqn.~\ref{eqn:rplmpl}. This near-cancellation of the gravity effect and the area effect is also seen in models of crust/lithosphere mass, e.g. \citealt{Kite2009}). For thicker magma shells, geometric corrections increase the value of the exponent. 

The magma configuration on sub-Neptunes - with a global magma layer directly in contact with the atmosphere - is similar to that on terrestrial planets immediately after a giant impact \citep{ElkinsTanton2012, Hamano2013}. The sub-Neptune magma configuration is very different from that on tidally-heated rocky worlds such as Io (which have a magma or silicate-mush layer under a solid rock layer). The sub-Neptune magma configuration is also very different from that on magma planets with silicate vapor atmospheres, which (in the absence of tidal heating) have a superficial ($<$100~km deep) dayside magma pool overlying solid rock \citep{Leger2011,Kite2016}. Sub-Neptune magma oceans are intrinsically much more abundant than magma planets \citep{Hsu2019,Winn2018}, long-lived (e.g. \citealt{Vazan2018a}), and massive (Fig.~\ref{fig:howmuchmagma}). It is possible that sub-Neptunes contain most of the magma in the Universe.
 
\section*{Appendix~B: \\ How magma gets oxidized.}

\noindent Protoplanetary disks with solar composition are very reducing because the H/O ratio is so large that the equilibrium \emph{f}O$_2$ is about six orders of magnitude below that of the iron - w\"{u}stite buffer. FeO-bearing silicates and magnetite are thermodynamically stable only at such low temperatures ($\le$600 K for FeO-bearing silicates, ($\le$400 K for magnetite) that their formation may be kinetically inhibited \citep{Krot2000, Lewis2004, Grossman2012}

Nevertheless, copious FeO is found in rocky-planet silicates \citep{RighterOBrien2011}. How do silicates get oxidized? One possibility is net loss of H$_2$ from the planet (or from the planet's building blocks). The other is rearrangement of oxidizing power inside the planet, with a net transfer of reductants to the Fe-metal core leaving oxidants in the silicate mantle. These two mechanisms can work in concert \citep{Fischer2015}. It turns out that loss of H$_2$ from the planet is expected if the planet's mass includes a large contribution from components that grew to km-size or larger outside the H$_2$O-ice snowline. Rearrangement of reducing power inside the planet is expected if an Fe-metal core equilibrates with silicate at high temperature, as is likely on $\gtrsim$1~$M_\Earth$ planets. 

\vspace{0.05in}
\noindent\emph{Oxidation mechanism 1: Net loss of reducing power from the whole planet.}
\vspace{0.05in}

\noindent The reaction 

\begin{equation}
\mathrm{Fe_{(s\,or\,l)} + H_2O_{(l)}} \rightarrow \mathrm{FeO_{(s\,or\,l)} + H_{2(g)}}
\label{eqn:feh2o}
\end{equation}

\noindent occurred on Solar System planetesimals \citep{Rosenberg2001, Zolensky1989, Zolensky2008, CastilloRogezYoung2017}. Another route to Fe-oxidation is during
hydration of Fe-silicates (alongside Mg-silicates) at $T$~$<$~800~K. For example:

\begin{equation}
\mathrm{6Fe_2SiO_4 + 11H_2O \rightarrow 2Fe_3O_4  + 3Fe_2Si_2O_5(OH)_4 + 5H_2}
\label{eqn:srp}
\end{equation}

\noindent  (``serpentinization'')  \citep{Sleep2004, McCollomBach2009, Klein2013}. 
For $R$~$<$~10$^{3}$~km objects, the gravitational binding energy for H$_2$ molecules is not much greater than their thermal energy. Thus, Reactions~\ref{eqn:feh2o}-\ref{eqn:srp} on planetesimals and small embryos cause return of  H$_2$ to the nebula (e.g., \citealt{LeGuillou2015}), and rock oxidation   \citep{Wilson1999, Rosenberg2001, Brearley2006}. Planets that form by collision between such pre-oxidized chunks can themselves be very oxidizing. 


Indeed, meteorites are much more oxidizing than the nebula \citep{Doyle2019}. Meteorite-based models of outgassing output high-molecular-weight atmospheres \citep{SchaeferFegley2010}. Solar System rocky planet H$_2$/H$_2$O ratios are $\ll$ 1 and correspond to log \emph{f}O$_2$ between approximately the iron (Fe) - w\"{u}stite (FeO) oxygen fugacity buffer (the ``IW buffer'') and the more-oxidizing quartz (SiO$_2$) - fayalite (Fe$_2$SiO$_4$) - magnetite (Fe$_3$O$_4$) oxygen fugacity buffer (the ``QFM buffer''). So, redox disequilibrium between nebula and magma, with the magma being more oxidized, is widespread in the Solar System.

We do not know how widespread these processes are in the Galaxy. Oxidation states inferred for extrasolar planetesimals based on white-dwarf data fall within the range for solar system silicates \citep{Doyle2019}. Generation of H$_2$ on diameter $<$~10$^3$~km bodies in the Solar System via Reaction~\ref{eqn:feh2o} is aided by $^{26}$Al decay. To the extent that the Solar System acquired a high dose of  $^{26}$Al, the Solar System might be a biased sample (\citealt{Lichtenberg2019}, but see also \citealt{Young2019}). 

Neither Reaction~\ref{eqn:feh2o} nor Reaction~\ref{eqn:srp} can operate on pebbles, because liquid water requires pressures $>$600 Pa (the triple point of H$_2$O) which is much more than the pressure in a pebble. Planets that grow by pebble accretion can generate H$_2$ by Reaction~\ref{eqn:feh2o}, but if they are large enough to retain the H$_2$ generated by these reactions, they might end up very reducing.

The connection between pebble accretion and sub-Neptune atmospheric composition that we are proposing only works if pebbles are not fragments of much larger bodies. Suppose instead that pebbles are dominantly debris from collisions between planetesimals. In this case, Reaction~\ref{eqn:feh2o} nor Reaction~\ref{eqn:srp} could have taken place within planetesimals that were then broken up to make pebbles.

\emph{Kepler} sub-Neptunes are inside the H$_2$O snowline. Whole-planet oxidation involves a contribution to the planet from beyond the H$_2$O snowline. This could involve planet migration  (e.g. \citealt{Cossou2014,Ogihara2015,Izidoro2017,Raymond2018,KiteFord2018, Carrera2019}). Alternatively, an oxidized sub-Neptune core could be assembled at its current location from planetesimals that migrated.

\vspace{0.05in}
\noindent\emph{Oxidation mechanism 2: Rearrangement of reducing power inside the planet.}
\vspace{0.05in}

\noindent At high temperature, Si can dissolve into liquid metal at wt\% levels via the reaction

\begin{equation}
\mathrm{SiO_{2 (silicate)} + 2Fe_{(metal)} \rightarrow Si_{(metal)} + 2 FeO_{(silicate)}}
\label{eqn:javoy}
\end{equation}

\noindent \citep{Javoy1995, Fischer2015}. Reaction~\ref{eqn:javoy} oxidizes the mantle. It is effective at oxidizing the mantle if fewer then two O atoms dissolve into the Fe-metal for every Si atom that dissolves into the Fe-metal. This condition is satisfied when the mantle is reduced. (This redox dependence suggests that reactions at the core-mantle boundary might be sensitive to redox reactions at the magma-atmosphere interface). Reaction~\ref{eqn:javoy} can account for at least half of the Earth's FeO content, although probably not all \citep{Rubie2011, Fischer2015}. 

On sub-Neptunes, Reaction~\ref{eqn:javoy} sugggests that magma oceans will have non-negligible volatile-free FeO content \citep{Wordsworth2018}. The only requirement is that liquid metal and magma chemically interact during accretion. 

Which of these oxidation mechanisms is more important? The FeO of Earth is less than that of Mars, even though Earth is bigger. This matches expectations for oxidation mechanism 1, but not oxidation mechanism 2. However, we have no idea whether or not this applies to exoplanets.

\begin{figure}
\centering
\includegraphics[width=1.1\columnwidth,clip=true,trim={38mm 0mm 0mm 0mm}]{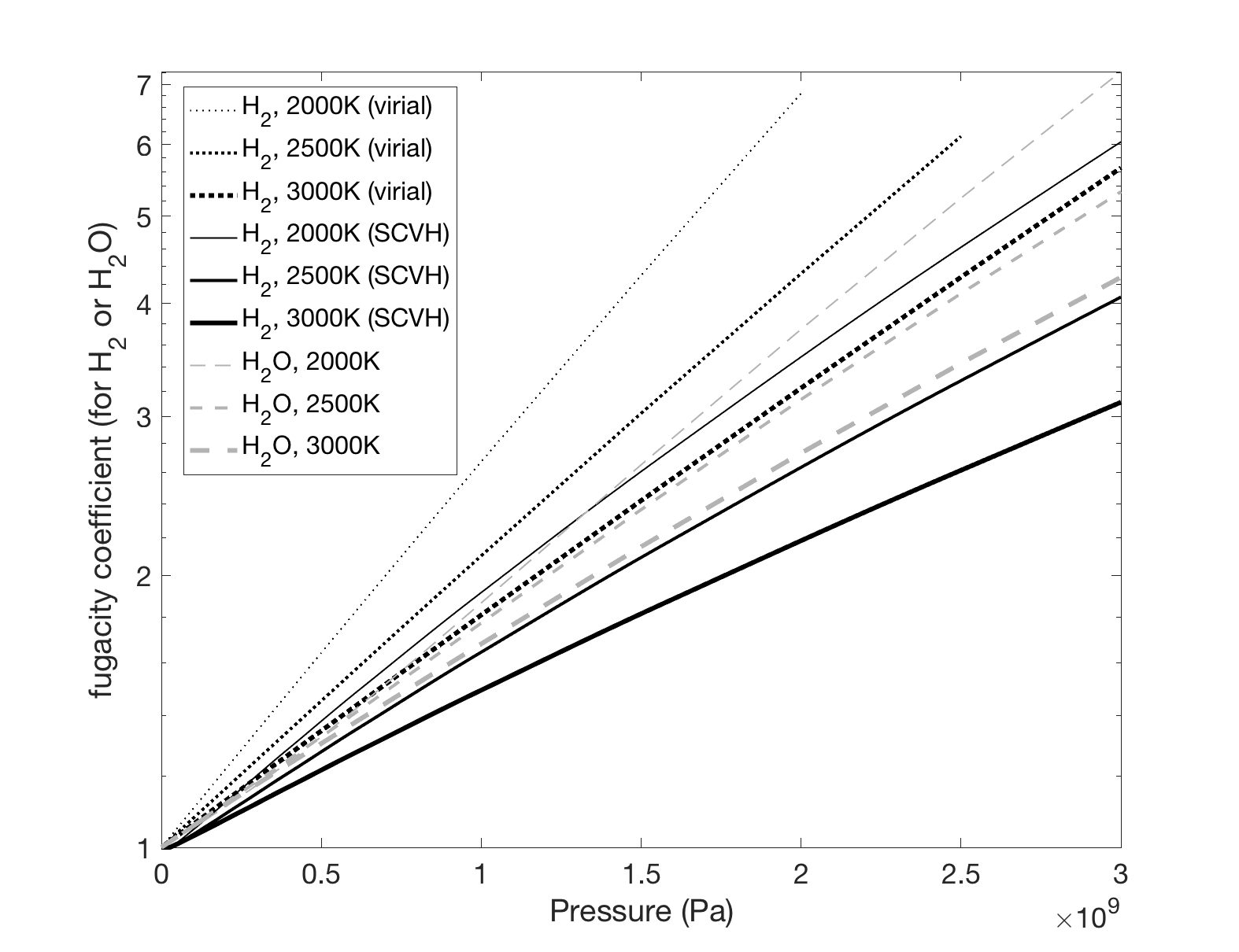}
\caption{Fugacity coefficients. For an ideal gas, the concentration of volatiles in the magma is proportional to the partial pressure of that volatile in the atmosphere above to the magma. The fugacity coefficient corrects for non-ideality of the gas and gives the effective pressure of real gases for solubility in melts and other chemical equilibria. ``SCVH'' refers to the hydrogen equation of state of \citet{Saumon1995}. See Appendix~C for details on how these fugacity coefficients were calculated.} 
\label{fig:fugacities}
\end{figure}

\section*{Appendix C: \\ Fugacity coefficients.} 

\noindent Our model includes non-ideal behavior of H$_2$ and of H$_2$O in the atmosphere.  Non-ideal behavior increases the solubility of sub-Neptune atmospheric components in magma (relative to the ideal case). Although we include non-ideal behavior of H$_2$ and of H$_2$O, we assume that H$_2$ and H$_2$O mix ideally. We calculated fugacity coefficients for the range 2000K-3000K and 0.01-2~GPa. We computed fugacity coefficients using both a virial equation of state and the \citet{Saumon1995} equation of state for H$_2$, and the \citet{Haar1984} EOS based on the Helmholtz energy for H$_2$O.

The fugacity coefficient of a gas is related to the the compressibility factor $Z$ by the equation

\begin{equation}
\mathrm{ln}\phi= \mathrm{ln} \frac{f}{P}= \int_0^P \left(\frac{Z-1}{P}\right) dP
\label{eqn:fugacityfromz}
\end{equation}

\noindent where $Z$ is given by

\begin{equation}
Z= \frac{PV_m}{RT}
\end{equation}

\noindent where $V_m$ is the molar volume.

We computed H$_2$ fugacity coefficients using a virial equation of state computed for an exponential repulsive potential \citep{AmdurMason1958, MasonVanderslice1958}.  Only the repulsive part of the potential is important at high-$T$ dense conditions, as discussed by \citet{AmdurMason1958}. The exponential repulsive potential is

\begin{equation}
\psi(r)=\psi_o \mathrm{exp}(-rÚ\rho)
\end{equation}

The values used for H$_2$ gas are $\psi_0$~=~10$^4$~eV and $\rho$~= 0.204$\times$10$^{-8}$~cm \citep{Bainbridge1962}. The asymptotic limiting equations for estimating $B$, $C$, and $D$ and the input to them are (from \citealt{HendersonOden1966}):

\begin{equation}
\frac{B(T)}{b_o} =(\mathrm{ln}\,\gamma x )^3
\end{equation}

\begin{equation}
\frac{C(T)}{b_o^2} =(\mathrm{ln}\,\gamma x )^6
\end{equation}

\begin{equation}
\frac{D(T)}{b_o^3} =(\mathrm{ln}\,\gamma x )^9
\end{equation}

where
$\gamma$ is Euler's constant and 

\begin{equation}
x=\psi_o/kT
\end{equation}

\begin{equation}
\gamma x=206,684,901/T
\end{equation}

\begin{equation}
b_o= \left(\frac{2\pi N_A}{3}\right) \rho^3=0.0107075
\label{eqn:b0}
\end{equation}

Here, $N_A$ is Avogadro's number.

While this approach gives good results for the range of magma-atmosphere interface conditions considered in this paper, it fails to converge above a limit roughly 0.3 of the H$_2$ molar volume at its critical point, i.e. about 3$\times$ the molar density at the critical point.

We also calculated fugacity coefficients from the H$_2$ EOS tables of of \citet{Saumon1995}. In this calculation, we forced $Z$~=~1 for $P$~$<$~10$^7$~Pa (the region of significant H$_2$ dissociation). Using the H$_2$ EOS tables of  \citet{Saumon1995} led to lower fugacity coefficients than the virial approach (Fig.~\ref{fig:fugacities}), implying lower importance for magma-atmosphere interactions than with the virial approach. To be conservative, we used the \citet{Saumon1995}-derived H$_2$ fugacities for this paper. Our conclusions in this paper are not qualitatively affected by this choice.

We computed H$_2$O fugacity coefficients from the \citet{Haar1984} EOS for water (the HGK EOS) as coded by \citet{Bakker2012}. \citet{Haar1984} developed an EOS based on the Helmholtz free energy of H$_2$O, which is valid to 2500~K and and 3~GPa. However, we used the Haar et al EOS over the entire $P$ - $T$ range studied because comparisons with literature results at high $P$ and $T$ were reasonable \citep{RiceWalsh1957, DelanyHelgeson1978, OstrovskyRizhenko1978}. For example, the HGK EOS gives molar volumes within 17\% of those reported by \citet{RiceWalsh1957} from 0.5~-~7~GPa on the 1273 K isotherm - the highest-T isotherm for which they present data. However, \citet{WagnerPruss2002} note Hugoniot data are not as precise as the other $P-V-T$ data used to derive the equation of state and they did not use the data as inputs to improve the EOS at very high $P$ and $T$.

Table 1 shows the H$_2$ virial coefficients used in this study. Fig.~\ref{fig:fugacities} shows the H$_2$ and H$_2$O fugacity coefficients used in this study. 

%
%

\begin{table}
\begin{center}
\caption{H$_2$ virial coefficients used to compute H$_2$ fugacity coefficients (Appendix C).  \label{tbl-999}}
\setlength\tabcolsep{3pt}
\begin{tabular}{lcccccccc}
\tableline\tableline
T/K &	  $B$(H$_2$) &			$C$(H$_2$) & $D$(H$_2$)   \\
\tableline\tableline
2000	& 	16.5 &		272  & 1285 \\
2500	& 	15.5	 &		242  & 1078 \\
3000	& 	14.8 &		219 & 930 \\
\tableline\tableline
\end{tabular}
\end{center}
\end{table}

\section*{Appendix~D: \\ Details of workflow.} \noindent Our \S3 workflow has four steps. Here we step through the adding-H$_2$ case (the adding-H$_2$O workflow is conceptually similar).  (Step~1)~We start from the initial mantle FeO content (before any H$_2$ has been added). From that, we find the \emph{f}O$_2$ (Fig.~\ref{fig:fo2}). Then, we find the H$_2$/H$_2$O ratio of fugacities from Eqns. \ref{eqn:h2o2}-\ref{eqn:deltagh2o2}. We convert this to partial pressures using the fugacity coefficients in Appendix~C. From the ratio of partial pressures plus the (imposed) total atmospheric pressure, we find the total mass of H$_2$ and of H$_2$O in the atmosphere. (Step~2)~We~then use the solubilities to figure the amounts of H$_2$ and of H$_2$O in the magma. We assume the magma is fully molten. We set an arbitrary upper limit of 25~wt\% H$_2$O in the magma (at this upper limit the melt is mostly H$_2$O in terms of mole fraction). (Step~3)~Equipped with the total H$_2$O in the system, plus the assumption that only H$_2$ is initially added, the amount of FeO reduction can be figured by mass balance. Making 1 mole of H$_2$O requires reduction of 1 mole of FeO. Thus the activity of ``FeO" and the \emph{f}O$_2$ are both lower than the initial values for H$_2$ addition. For mass balance purposes, we assume an activity coefficient for FeO of 1.0 (we use an FeO activity coefficient of 1.50 when calculating the \emph{f}O$_2$ ). We do not consider the possible effect of volatiles on the buffer equilibria \citep{Bezmen2016}. (Step~4)~To make the calculation self consistent, we iterate steps 1-3. We plug in the new value of \emph{f}O$_2$. We find the new amount of total H$_2$O produced (which will be lower). We recalculate the FeO decrement until convergence.  
 

 The Bulk Silicate Earth composition of \citet{SchaeferFegley2009} is used for mass balance, except for  the extreme redox endmembers, where we follow the compositions of \citet{ElkinsTantonSeager2008b}.

\noindent 

\section*{Appendix~E: \\ Solar System Connections.} \noindent Volatile-rock reactions have long been considered as potential sources of H$_2$O and H$_2$ on Earth and Mars. \citet{Atreya1989} provides a comprehensive review, and more recent work is summarized in \citet{DauphasMorbidelli2014}. For example, the Fe$^{2+}$ in Earth rocks and Mars rocks probably derives, at least in part, from accretion of building blocks that underwent the net reaction Fe + H$_2$O$\rightarrow$ FeO + H$_2$ (e.g. \citealt{Ringwood1979, Rubie2015}) (Appendix~B). On the other hand, for~Earth and Mars, nebula gas is no longer thought to have contributed much to the mass of the present-day atmospheres and oceans (e.g.,~\citealt{DauphasMorbidelli2014}; but see also \citealt{OlsonSharp2019}). These Earth and Mars studies did not enforce redox balance; the H$_2$ must have almost entirely escaped for Earth and Mars. However this cannot be true for sub-Neptunes.  

Solar System readers should be aware that, in papers about sub-Neptunes, the ``core'' consists of both Fe-metal and silicates; ``rocky'' can refer to either solid rock or liquid silicate (magma); ``rocky planet'' is generally taken to exclude sub-Neptunes even though Kepler sub-Neptunes are mostly silicates by mass; the terms ``sub-Neptune'' and ``mini-Neptune'' are used interchangeably; and the terms ``envelope'' and ``atmosphere'' are also used interchangeably. For sub-Neptunes the atmosphere ``gas'' is mostly supercritical fluid.


\clearpage

\end{document}